\documentclass[final,3p,times]{elsarticle}

\usepackage{geometry}
\usepackage{amssymb}

\usepackage{amsmath}
\usepackage{tensor}
\usepackage{slashed}
\usepackage{amsfonts}
\usepackage{amssymb}
\usepackage{bm}
\usepackage{dsfont}
\usepackage{multirow}
\usepackage{ulem}

\usepackage{hyperref}
\hypersetup{
     colorlinks   = true,
     citecolor    = blue,
     urlcolor = blue,
     linkcolor = blue
}

\makeatletter
\newcommand{\xleftrightarrow}[2][]{\ext@arrow 3359\leftrightarrowfill@{#1}{#2}}
\makeatother
\newcommand{\beq}{\begin{equation}}
\newcommand{\eeq}{\end{equation}}
\newcommand{\beqa}{\begin{eqnarray}}
\newcommand{\eeqa}{\end{eqnarray}}

\journal{Annals of Physics}

\begin{document}

\begin{frontmatter}



\title{Boson-fermion duality in a gravitational background}


\author[a]{Yago Ferreiros}
\author[b]{Eduardo Fradkin}

\address[a]{Department of Physics, KTH-Royal Institute of Technology,\\SE-106 91 Stockholm, Sweden}
\address[b]{Institute for Condensed Matter Theory and Department of Physics,
\\University of Illinois at Urbana-Champaign, 1110 West Green St, Urbana Illinois 61801-3080, U.S.A.}

\begin{abstract}
We study the 2+1 dimensional boson-fermion duality in the presence of background curvature and electromagnetic fields. The main players are, on the one hand, a massive complex $|\phi|^4$ scalar field coupled to a $U(1)$ Maxwell-Chern-Simons gauge field at level $1$, representing a relativistic composite boson with one unit of attached flux, and on the other hand, a massive Dirac fermion. We show that, in a curved background and at the level of the partition function, the relativistic composite boson, in the infinite coupling limit, is dual to a short-range interacting Dirac fermion. The coupling to the gravitational spin connection arises naturally from the spin factors of the Wilson loop in the Chern-Simons theory. A non-minimal coupling to the scalar curvature is included on the bosonic side in order to obtain agreement between partition functions. Although an explicit Lagrangian expression for the fermionic interactions is not obtained, their short-range nature constrains them to be irrelevant, which protects the duality in its strong interpretation as an exact mapping at the IR fixed point between a Wilson-Fisher-Chern-Simons complex scalar and a free Dirac fermion. We also show that, even away from the IR, keeping the $|\phi|^4$ term is of key importance as it provides the short-range bosonic interactions necessary to prevent intersections of worldlines in the path integral, thus forbidding unknotting of knots and ensuring preservation of the worldline topologies.
\end{abstract}


\end{frontmatter}


\section{Introduction}

Boson-fermion dualities have been investigated for a long time. The earliest version of this type of duality is the bosonization of fermions in 1+1 space-time dimensions \cite{ML65,LE74,C75,M75}. This is a mapping between a massless Dirac (spinor) fermion $\Psi$ and a massless compactified boson $\varphi$, with compactification radius $R=2\pi$,
\beq
i\bar{\psi}(\slashed{\partial}+i\slashed{A})\psi\longleftrightarrow \frac{1}{8\pi}(\partial_\mu\varphi)^2-\frac{1}{2\pi}\epsilon^{\mu\nu}A_\mu\partial_\nu \varphi.
\label{eq:bosonization-1+1}
\eeq
$A_\mu$ is an external background gauge field, and $\slashed{\partial}=\gamma^\mu\partial_\mu$, with $\gamma^\mu$ the gamma matrices. The fermionic current maps to
\beq
j^\mu=\bar{\psi}\gamma^\mu\psi\longleftrightarrow\frac{1}{2\pi}\epsilon^{\mu\nu}\partial_\nu \varphi.
\label{eq:dual-j-1+1}
\eeq
The 1+1 dimensional boson-fermion duality  is a mapping at the operator level, e.g. the fermion mass term ${\bar \psi} \psi$ maps onto the vertex operator $\cos \varphi$. It is  also at the level of the spectrum and at the level of the partition function. 

Things get a bit more subtle in 2+1 dimensions. The first hints at the existence of a 2+1 dimensional boson-fermion duality were found when it was discovered that attaching fluxes to particles could transmute their statistics, becoming anyons \cite{W82,WZ83,ASWZ85,HRZ88}, a concept that has wide and deep applications particularly in the context of the fractional quantum Hall effects \cite{J89,ZHK89,LF91}. So if one attaches an odd number of units of flux to a boson, one obtains a composite boson with fermionic statistics. Because  these early works dealt with non relativistic composite particles, one may  wonder how does  flux attachment work in the relativistic case. This is a far more subtle question, because if a relativistic boson, e. g. a scalar field, acquires fermionic statistics under the attachment of an odd number of units of flux, it would give rise to a relativistic fermion. Relativistic fermions are spinors, so now one would have to understand how a relativistic scalar field transmutes into a relativistic spinor Fermi field.

It was Polyakov who first proposed the boson-fermion duality for relativistic massive fields in 2+1 dimensions \cite{P88}. He showed that the propagator of a massive free complex scalar field, coupled to a dynamical $U(1)$ Chern-Simons (CS) field \cite{DJT82,W88} at level 1, maps to the propagator of a massive Dirac fermion. The CS field is responsible for the flux attachment, the level 1 corresponding to one unit of flux. Polyakov's duality reads\footnote{Polyakov's mapping \cite{P88} actually involved a doublet of complex scalars; we will see that a single complex scalar field suffices.}
\beq
i\bar{\psi}\slashed{\partial}\psi+m\bar{\psi}\psi\longleftrightarrow |(\partial_\mu+ia_\mu)\phi|^2-m^2|\phi|^2+\frac{1}{4\pi}\epsilon^{\mu\rho\nu}a_\mu\partial_\rho a_\nu.
\label{eq:duality-intro}
\eeq
Polyakov's result  inspired much subsequent work, most of which focused on the role of different regularization schemes and the associated problem of the framing of knots \cite{T88,AS88,GH89,H89,CL89,IIM901,IIM90,SSS90,SS91}. However, aspects of Polyakov's mapping are puzzling. First, if one tries to extend the duality of Eq. \eqref{eq:duality-intro} by minimally coupling the theories at both sides to a background U(1) gauge field and curvature, one finds that Eq. \eqref{eq:duality-intro} maps theories with different anomalies: on one side we have a Dirac fermion with a parity anomaly \cite{R84,GPM85,GV86,V86,GRS96,KV18}, and on the other side a CS gauge field with a framing anomaly \cite{W89,BW91}. The existence of non matching anomalies presents an obstruction for the duality to be realized, as it stands, as an exact mapping between partition functions\footnote{The original work of Polyakov was done in the absence of background fields, so the difficulty of having anomalies was not apparent.}. Second, the duality relies on a topological property of Wilson loops in the CS theory: closed worldline configurations are classified in terms of a topological invariant called Gauss's linking number \cite{G77}, which counts the number of loops of one worldline around another. If the scalar field is non interacting, then there is no interaction between worldlines and nothing would prevent them to be unknotted, so the topology of the configurations of worldlines would not be preserved. And third, Polyakov's duality is seemingly valid only for massive free scalar fields. In particular, it cannot be valid in the massless theory whose behavior is controlled by the Wilson-Fisher fixed point. This has come back to focus with the recent works on conjectured CS-matter-dualities in 2+1 dimensions \cite{AGY12,JMY13,SW16,KT16,A16,HS16,NNB16,KRT17,ABHS17,MVX17,JK17,BHS17,CHS217,CHS17,ABKR17,B18,J17,JBY18,CZ18}. Notice that the field content of Polyakov's 2+1-dimensional bosonization duality, Eq.\eqref{eq:duality-intro}, and the 1+1-dimensional bosonization duality, Eq.\eqref{eq:bosonization-1+1}, are quite different: the 1+1-dimensional scalar field $\varphi$ is real and compact and should be regarded as a (failed) Goldstone boson, while Polyakov's is a non-compact complex scalar field $\phi$. In this sense the analogy is between the 2+1 dimensional complex boson and the vertex operators of the comapctified 1+1 dimensional real boson.

A different perspective on 2+1-dimensional dualities was introduced in the work of Fradkin and Kivelson \cite{FK96} who introduced a loop model to describe the quantum fluctuations of a system with flux-attachment. In this model, originally introduced as a model of a putative quantum critical point for the quantum Hall plateau transitions,  the quantum theory is represented by the closed worldlines of particles and holes (antiparticles).  Key to this construction is the assumption that the loops cannot intersect, which always holds for interacting scalar fields below four space-time dimensions. Flux attachment was implemented by demanding that the action of a  configuration of loops has an imaginary part proportional to the linking number of the configuration (which is a stable concept if the loops cannot intersect). By imposing that the parity-even part of the action scales in the same way as the imaginary part, Ref. \cite{FK96} showed that  the partition function is invariant under an extended duality in the absence of background gauge fields, and it is covariant under the extended duality in the presence of background gauge fields.

Ref. \cite{FK96} ignored the self-linking of the worldlines, which led to the existence of an exact invariance under periodic changes of the statistical angle and a full $SL(2,\mathbb{Z})$ exact symmetry. In a recent and separate paper Goldman and one of us \cite{GF18} has reexamined the issue of the existence of an exact periodicity in  asymptotically massless models with Lorentz invariance, and showed that the duality can be implemented consistently, although the existence of fractional spin is generic and, hence, exact periodicity does not hold. Therefore, although loop models can be regarded as a UV completion of the conjectured CS-matter-dualities, extended $SL(2,\mathbb{Z})$ invariance is not generally allowed as an exact symmetry of a partition function. Ref. \cite{GF18} showed that loop representations can be used to make Polyakov's duality consistent with the parity anomaly for background electromagnetic fields, and  used loop models to check the conjectured web of dualities \cite{KT16,SW16}.

On a separate track, in the 1990s other versions of bosonization in 2+1 dimensional  relativistic theories were also introduced \cite{BQ93,FS94,CLQ94,LMSN97}. These theories use the parity anomaly to map the effective low energy action of conserved currents of theories of massive Dirac fermions to a U(1) Chern-Simons gauge theory. The bosonization rule reads
\begin{equation}
j_\mu \leftrightarrow \frac{1}{2\pi} \epsilon_{\mu \nu \lambda} \partial^\nu a^\lambda
\label{eq:dual-j-2+1}
\end{equation}
where $j_\mu$ is the fermionic current and $a_\mu$ is the Chern-Simons gauge field. Notice that this equation has a  similar form to Eq. \eqref{eq:dual-j-1+1}. In particular, it relates the fermionic density  to the gauge flux. In this sense it is an electric-magnetic duality, and resembles  the particle-vortex duality of Refs. \cite{P78,TS78,DH81}. This fermion-gauge duality was recently used to derive an effective hydrodynamic gauge theory of topological insulators \cite{CJRF13,CKRF16}.


Recently, largely inspired on previous results valid in the large-$N$ limit of fermionic and bosonic theories coupled to Chern-Simon gauge fields \cite{AGY12,JMY13}, it was conjectured \cite{KT16,SW16} that the IR version of the boson-fermion duality maps a massless Dirac fermion to a (strongly) interacting gauged complex scalar field at its Wilson-Fisher fixed point, coupled to a level 1 Chern-Simons gauge field,
\beq
i\bar{\psi}(\slashed{\partial}+i\slashed{A})\psi\longleftrightarrow |(\partial_\mu+ia_\mu)\phi|^2-\lambda|\phi|^4+\frac{1}{4\pi}\epsilon^{\mu\rho\nu}a_\mu\partial_\rho a_\nu+\frac{1}{2\pi}\epsilon^{\mu\rho\nu}a_\mu\partial_\rho A_\nu.
\label{eq:wilson-fisher-intro}
\eeq
The validity of this mapping  relies on the identification of operators on both sides. In particular, the fermionic current of the left hand side is identified with the gauge flux of the right hand side, in agreement with Eq. \eqref{eq:dual-j-2+1}. Very recently, a proof of this duality at the partition function level has been achieved by using an exact UV correspondence on the lattice \cite{CSWR17}. The mapping of Eq. \eqref{eq:wilson-fisher-intro} applies in two distinct physical settings, each with different properties and symmetries. One is a theory of Dirac fermions in 2+1 dimensions. In that case, if the  theory is defined with  a gauge-invariant UV regularization, a time-reversal breaking counterterm of the form $\pm \frac{1}{8\pi} \epsilon_{\mu \nu \lambda} A^\mu \partial^\nu A^\lambda$  (the sign depends on the choice of regularization) should be added to the left hand side of the mapping of Eq. \eqref{eq:wilson-fisher-intro} to account for the time-reversal (or parity) anomaly \cite{R84,W16}. On the other hand, a manifestly time-reversal invariant theory is possible if the 2+1 dimensional theory is the boundary of a 3+1 dimensional bulk whose effective action is a U(1) gauge theory with a $\theta=\pi$ term, which cancels the parity (or time-reversal) anomaly of the boundary fermions \cite{SW16}.

A  comparison of Polyakov's duality, Eq. \eqref{eq:duality-intro}, with the duality of Eq. \eqref{eq:wilson-fisher-intro} suggests that they are not quite consistent with each other. To begin with, in Eq. \eqref{eq:duality-intro} the complex scalar field is free whereas in Eq. \eqref{eq:wilson-fisher-intro} it is interacting and tuned to the (non-trivial) Wilson-Fisher fixed point. Since the fermionic side of Eq. \eqref{eq:duality-intro} has a mass term it is more relevant to make the comparison with the massive version of Eq. \eqref{eq:wilson-fisher-intro}. In this case, it was noted in Refs. \cite{KT16,SW16} that the mapping depends on the sign of the fermionic mass. For one sign of the mass term and one choice of regularization, the fermionic theory is in a topological non-trivial phase and maps onto a massive interacting complex scalar in its unbroken (symmetric) phase. This case is consistent with Polyakov's. However for the opposite sign of the mass, the fermionic theory is trivial and the dual scalar theory is in its broken (Higgs) phase. As it stands, this case cannot be described by Polyakov's mapping. We will see that there is a way to reconcile these apparent discrepancies.

References \cite{KT16,SW16} used the duality of Eq. \eqref{eq:wilson-fisher-intro} as the building block to derive other mappings, arriving to a whole web of dualities in 2+1 dimensions. One of these derived dualities is a map between two Dirac fermions, which was first proposed in refs. \cite{S15,MV16,WS15}. This fermion-fermion duality reads
\beq
i\bar{\psi}(\slashed{\partial}+i\slashed{A})\psi\longleftrightarrow i\bar{\chi}(\slashed{\partial}+i\slashed{a})\chi+\frac{1}{4\pi}\epsilon^{\mu\rho\nu}a_\mu\partial_\rho A_\nu.
\label{eq fermion fermion duality}
\eeq
where the $\chi$ fermions on the right hand side (r.h.s) are {\it non-local} respect to the $\psi$ fermions of the left hand side (l.h.s) and act on different Hilbert spaces. The dual Dirac theory of the r.h.s. of Eq. (\ref{eq fermion fermion duality}) has been related \cite{MV16,M15,WS16} to the bulk electric-magnetic duality of a three dimensional topological insulator \cite{FKM07,QHZ08,HK10,QZ11}, while a derivation of the fermion-fermion duality from a wire construction has also appeared in the literature \cite{MAM16,MAM17}. This fermionic mapping has been used to construct a particle-hole symmetric theory of the half-filled Landau level \cite{S15}, and all known surface states of the three dimensional topological insulator have been also successfully accessed from the dual Dirac fermion \cite{WS15}.

In this work we re-examine Polyakov's duality as an exact mapping between partition functions, studying it's validity in the presence of background metric (curvature), and also gauge fields, and giving an answer to the questions and puzzles stated above. Apart from contributing to the fundamental understanding of the boson-fermion duality, our work opens the door to the study of gravitational responses of relativistic composite particles. For non relativistic composite particles, gravitational responses have been widely studied in the context of the fractional quantum Hall effect \cite{WZ92,H09,RR11,S13,AG14,GA14,CYF14,GA15,GF15}. These include thermal responses like the thermal Hall effect \cite{KF97,CHZ02,WQZ11,RML12,S12}, which can be related to responses to a gravitational potential by Luttinger's formalism \cite{L64,QNS11,BR15}, and geometric responses like the Hall viscosity \cite{ASZ95,R09,RR11,HS12,BGR12,H09,HLF11,NS11,HLP13,AG14,FHS14,CFLV16}, which are related to the way the fluid couples to geometric properties of the two-dimensional surface on which it moves.

The layout of the paper is as follows. In section \ref{sec:starting-point} we compute the effective action of a relativistic composite boson, in the presence of background curvature and electromagnetic field, and express it as a worldline path integral. In section \ref{sec:effactionfermions} we apply an analogous procedure to a Dirac fermion. In section \ref{sec:dualitytimereversal} we compare the effective actions derived in the previous two sections, and obtain the boson-fermion duality when the bosonic and fermionic theories are both in the trivial and the topological insulating phases. Section \ref{sec:conclusions} is devoted to the conclusions. In order to easing the reading of the manuscript, some derivations and technicalities have been left as appendices.

\section{Partition function for the relativistic composite boson}
\label{sec:starting-point}

Let us start with the three dimensional action for a complex $|\phi|^4$ scalar field in a curved background with Minkowski metric signature, minimally coupled to a $U(1)$ Maxwell-CS gauge field $a_\mu$ and to a background field $A_\mu$, with $\mu=1,2,3$, and with a 
conformal coupling to the Ricci scalar curvature $R$, 
\cite{Brown-1977,Dowker-1976,Deser-1976,Drummond-1979}
\begin{align}
S_{scalar+flux}=&\int d^3x \sqrt{g}\left[\big|D_\mu(a,A)\phi\big|^2-\Big(\frac{R}{4}+m^2\Big)|\phi|^2-\lambda|\phi|^4\right]+S_{GF} \nonumber\\
S_{GF}=&-\frac{1}{4e}\int d^3x \sqrt{g}\,F^2(a)+CS_a,
\label{eq action composite boson}
\end{align}
where $D_\mu(a,A)=\partial_\mu+ia_\mu+iA_\mu$, $|D_\mu\phi|^2=g^{\mu\nu}(D_\mu\phi)^*D_\nu\phi$, $F^2=g^{\mu\rho}g^{\nu\gamma}F_{\mu\nu}F_{\rho\gamma}$, and $F_{\mu\nu}(a)=\partial_\mu a_\nu-\partial_\nu a_\mu$. $CS_a$ is
\beq
CS_a=\frac{1}{4\delta}\int d^3x\epsilon^{\mu\rho\nu}a_\mu\partial_\rho a_\nu,
\eeq
where we choose the statistical angle to be $\delta=\pi$, so that we have a level 1 CS term. We included a Maxwell $F^2$ term for $a_\mu$ which, as we will see,  will serve  as an UV regulator. We can introduce an auxiliary scalar field $\sigma$ and write
\beq
S_{scalar+flux}=\int d^3x \sqrt{g}\left[\big|D_\mu(a,A)\phi\big|^2-\Big(\frac{R}{4}+m^2+2i\sqrt{\lambda}\,\sigma\Big)|\phi|^2-\sigma^2\right]+S_{GF},
\eeq
such that integrating it out we recover the action \eqref{eq action composite boson}.
The partition function reads
\beq
Z_{scalar+flux}=\int D\phi^*D\phi D\sigma Da_\mu\, e^{iS_{scalar+flux}}.
\eeq
Integrating out $\phi$ one obtains, after a Wick rotation $t\rightarrow-it$ and using the identity $\ln\det=\mathrm{tr}\ln$ \footnote{Now the notation reads $S_{GF}=\frac{1}{4e}\int d^3x \sqrt{g}\,F^2(a)-iCS_a$},
\beq
Z_{scalar+flux}=\int D\sigma Da_\mu\,e^{-\Gamma'}\,e^{-\int d^3x\,\sigma^2}\,e^{-S_{GF}},
\eeq
with the effective action $\Gamma'$
\beq
\Gamma'=\mathrm{tr}\ln\Big[-\frac{1}{\sqrt{g}}D_\mu(a,A)\sqrt{g}\,g^{\mu\nu}D_\nu(a,A)+\frac{R}{4}+m^2+2i\sqrt{\lambda}\,\sigma\Big].
\label{eq primed effective action}
\eeq
Now the metric $g^{\mu\nu}$ has euclidean signature. Following appendix \ref{sec:worldline}, $\Gamma'$ can be written as a path integral over worldlines
\beq
-\Gamma'=\int_{\epsilon}^\infty dt\,t^{-1}\,\int DX^\mu\,e^{-\int_0^tds\big(\frac{1}{4}\dot{X}^2-ia_\mu\dot{X}^\mu-iA_\mu\dot{X}^\mu+\frac{R}{4}+m^2+2i\sqrt{\lambda}\,\sigma\big)},
\label{Eq. effective action boson}
\eeq
with $X_\mu(t)=X_\mu(0)$ and where we introduced a cut-off $\epsilon$.

\subsection{Integrating out the Chern-Simons field}
Now we proceed to integrate out the CS field. We can write
$$
Z_{scalar+flux}=\int D\sigma Da_\mu\,e^{-\Gamma'}\,e^{-\int d^3x\,\sigma^2}\,e^{-S_{GF}}
$$
\beq
=Z_{GF}\int D\sigma\,e^{-\int d^3x\,\sigma^2}\left\langle \exp\Big[\int_{\epsilon_s}^\infty dt\,t^{-1}\,DX^\mu\,e^{-\int_0^tds\big(\frac{1}{4}\dot{X}^2-iA_\mu\dot{X}^\mu+\frac{R}{4}+m^2+2i\sqrt{\lambda}\,\sigma\big)} e^{i\int_0^tds\,\dot{X}^\mu a_\mu}\Big]\right\rangle_{CS},
\eeq
where 
\beq
Z_{GF}=\int Da_\mu\,e^{-S_{GF}},
\eeq
and $\langle...\rangle_{CS}$ denotes the expectation value over the CS field
\beq
\langle...\rangle_{CS}=Z_{GF}^{-1}\int Da_\mu\,(...)\,e^{-S_{GF}},
\eeq
We can write the exponential as a series
\begin{align}
&\left\langle\sum_{n=0}^\infty\frac{1}{n!}\Big(\int_{\epsilon_s}^\infty dt\,t^{-1}\,DX^\mu\,e^{-\int_0^tds\big(\frac{1}{4}\dot{X}^2-iA_\mu\dot{X}^\mu+\frac{R}{4}+m^2+2i\sqrt{\lambda}\,\sigma\big)} e^{i\int_0^tds\,\dot{X}^\mu a_\mu}\Big)^n\right\rangle_{CS} \nonumber\\ 
&=1+\sum_{n=1}^\infty\frac{1}{n!}\left(\prod_{i=1}^n\int_{\epsilon_s}^\infty dt_i\,t_i^{-1}\,DX^\mu_i\,e^{-\int_0^{t_i}ds\big(\frac{1}{4}\dot{X}^2_i-iA_\mu\dot{X}^\mu_i+\frac{R}{4}+m^2+2i\sqrt{\lambda}\,\sigma\big)}\right)\Big\langle\prod_{i=1}^n e^{i\int_0^{t_i}ds\,\dot{X}_i^\mu a_\mu}\Big\rangle_{CS}.
\end{align}
\begin{figure}
\centering
\includegraphics[scale=0.6]{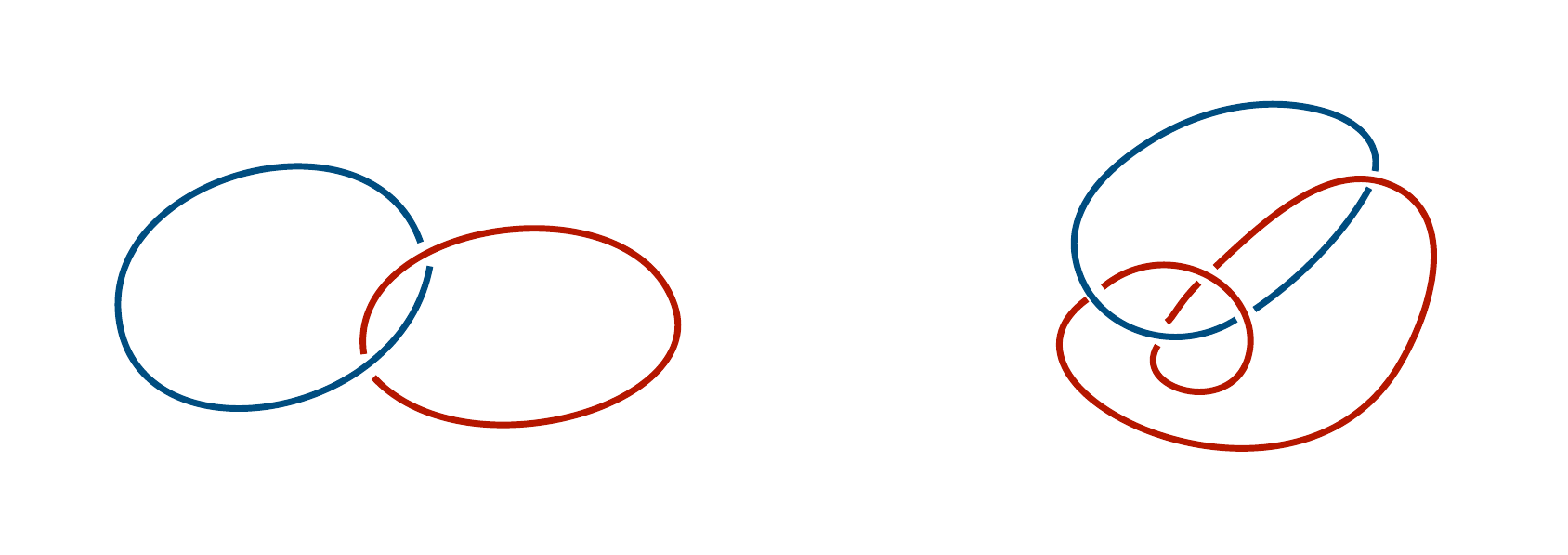}
\caption{Linking of two worldlines. The Gauss's linking numbers of the left and right configurations are $\pm1$ and $\pm2$, respectively (the sign depends on the relative orientation of the worldlines).}
\label{Fig:linkingnumber}
\end{figure}
Integrating out the scalar field $\sigma$ we get (up to a multiplicative constant)
\begin{align}
Z_{scalar+flux}\,Z_{GF}^{-1}=&1+\sum_{n=1}^\infty\frac{1}{n!}\left(\prod_{i=1}^n\int_{\epsilon_s}^\infty dt_i\,t_i^{-1}\,DX^\mu_i\,e^{-\int_0^{t_i}ds\big(\frac{1}{4}\dot{X}^2_i-iA_\mu\dot{X}^\mu_i+\frac{R}{4}+m^2\big)}\right)
\nonumber\\
&\times\left(\prod_{i,j=1}^n e^{-\lambda\int_0^{t_i}ds\int_0^{t_j}ds'\,\delta(X_i-X_j)}\right)\,\Big\langle\prod_{i=1}^n e^{i\int_0^{t_i}ds\,\dot{X}_i^\mu a_\mu}\Big\rangle_{CS}.
\end{align}
Hence we obtain the well known result that the $\phi^4$ interaction amounts to a short range repulsive interaction between worldlines \cite{Symanzik1969,Aizenman1981,Frohlich-1982}. Such a term suppresses intersecting worldline configurations, which is key as we will  see below.

We can now proceed to compute the expectation value of the product of Wilson loops (in the Feynman gauge) \cite{P88,H89}
\begin{align}
&\Big\langle\prod_{i=1}^n e^{i\int_0^{t_i}ds\,\dot{X}_i^\mu a_\mu}\Big\rangle_{CS}=\Big\langle\prod_{i=1}^n e^{i\oint_{C_i} dx^\mu a_\mu}\Big\rangle \nonumber\\
&=\prod_{i,j=1}^n \exp\left\{\frac{e^2}{8\pi}\oint_{C_i} dx^\mu\oint_{C_j} dy^\nu\Big[i\epsilon_{\mu\nu\rho}\frac{r^\rho}{\mu r^3}\big(1-(1+\mu r)e^{-\mu r}\big)-\delta_{\mu\nu}\frac{e^{-\mu r}}{r}\Big]\right\},
\end{align}
where $r=|x-y|$, and $\mu=e^2/2\delta$ (in our case $\delta=\pi$) has the effect of screening the interactions between worldlines, down to the non-local linking number. Here the Maxwell term acts as an UV regulator, with coupling constant $e$. If $C_i=C_j$ we get the expectation value of a single Wilson loop
\beq
\exp\left\{\frac{e^2}{8\pi}\oint_{C_i} dx^\mu\oint_{C_i} dy^\nu\Big[i\epsilon_{\mu\nu\rho}\frac{r^\rho}{\mu r^3}\big(1-(1+\mu r)e^{-\mu r}\big)-\delta_{\mu\nu}\frac{e^{-\mu r}}{r}\Big]\right\}\equiv\Big\langle e^{i\oint_{C_i} dx^\mu a_\mu}\Big\rangle_{CS},
\label{eq single Wilson loop}
\eeq
which we will look at more carefully in the next two sections. For the case of distinct closed loops $C_i$, $C_j$, and in the limit $e\rightarrow\infty$ we get 
\beq
\exp\left[\frac{1}{4}\oint_{C_i} dx^\mu\oint_{C_j} dy^\nu i\epsilon_{\mu\nu\rho}\frac{r^\rho}{r^3}\right]=\pi L(C_i,C_j).
\label{eq integral linking number}
\eeq
$L(C_i,C_j)$ is the Gauss linking number of $C_i$ and $C_j$ \cite{G77}. It is a topological invariant which takes integer values (see figure \ref{Fig:linkingnumber}). However, $L(C_i,C_j)$ is well defined only if $C_i$ and $C_j$ do not intersect. But we have seen that the $|\phi|^4$ self-interaction provides Dirac delta type interactions between worldlines, such that configurations with intersections are suppressed. Or in other words, the $|\phi|^4$ term is key to prevent links and knots from being unlinked and unknotted, preserving their topology.
We then have
\begin{align}
\Big\langle\prod_{i=1}^n e^{i\oint_{C_i} dx^\mu a_\mu}\Big\rangle_{CS}=\left(\prod_{i=1}^n\Big\langle e^{i\oint_{C_i} dx^\mu a_\mu}\Big\rangle_{CS}\right) \exp\Big(i\pi\sum_{i<j}2L(C_i,C_j)\Big)\nonumber\\
&=\prod_{i=1}^n\Big\langle e^{i\oint_{C_i} dx^\mu a_\mu}\Big\rangle_{e\rightarrow\infty},
\end{align}
where we took the limit $e \to \infty$ (where the Maxwell term is formally absent) and
where, in the summation over $i,j$, intersecting curves are excluded. In these expressions the remaining expectation values of the gauge field involve only one loop at a time. We will show below that these factors play an important role.

We see then that the expectation value of the product of Wilson loops factorizes. It is important to note that this is only the case if the statistical angle is $\delta=n\pi$, $n\in\mathbb{Z}$. For other values of $\delta$, the exponential $\exp(i\delta\sum_{i<j}2L)$ would induce interactions between Wilson loops. For the partition function we get, in the limit $e\rightarrow\infty$,
\begin{align}
Z^{e\rightarrow\infty}_{scalar+flux}\,Z_{CS}^{-1}&=\\ \int D\sigma\,e^{-\int d^3x\,\sigma^2}
& \exp\Big[\int_{\epsilon_s}^\infty dt\,t^{-1}\,DX^\mu\,e^{-\int_0^tds\big(\frac{1}{4}\dot{X}^2-iA_\mu\dot{X}^\mu+\frac{R}{4}+m^2+2i\sqrt{\lambda}\,\sigma\big)} \Big\langle e^{i\int_0^tds\,\dot{X}^\mu a_\mu}\Big\rangle\Big],
\end{align}
where we have taken the limit\footnote{Actually, there is a subtlety in doing the limit $e\rightarrow\infty$ in $Z_{GF}$. See comment in section \ref{sec:framing-anomaly}.} $Z^{e\rightarrow\infty}_{GF}=Z_{CS}$, with $Z_{CS}$ being  the partition function for the pure CS theory, and where we  restored the functional integration over $\sigma$. We will express these results  as a path integral over the $\sigma$ fields,
\beq
Z^{e\rightarrow\infty}_{scalar+flux}=Z_{CS}\int D\sigma\,e^{-\int d^3x\,\sigma^2}\,e^{-\Gamma_{scalar+flux}},
\label{eq z scalar+flux}
\eeq
where the effective action $\Gamma_{scalar+flux}$ is
\beq
-\Gamma_{scalar+flux}=\int_{\epsilon}^\infty dt\,t^{-1}\,\int DX^\mu\,e^{-\int_0^tds\big(\frac{1}{4}\dot{X}^2-iA_\mu\dot{X}^\mu+\frac{R}{4}+m^2+2i\sqrt{\lambda}\,\sigma\big)}\, \Phi(C),
\label{eq path integral boson}
\eeq
with $\Phi(C)$ the expectation value of the Wilson loop
\beq
\Phi(C)=\big\langle e^{i\oint_{C} dx^\mu a_\mu}\big\rangle_{e\rightarrow\infty}.
\eeq

\subsection{The Wilson loop in flat space-time}

\begin{figure}
\centering
\includegraphics[scale=0.65]{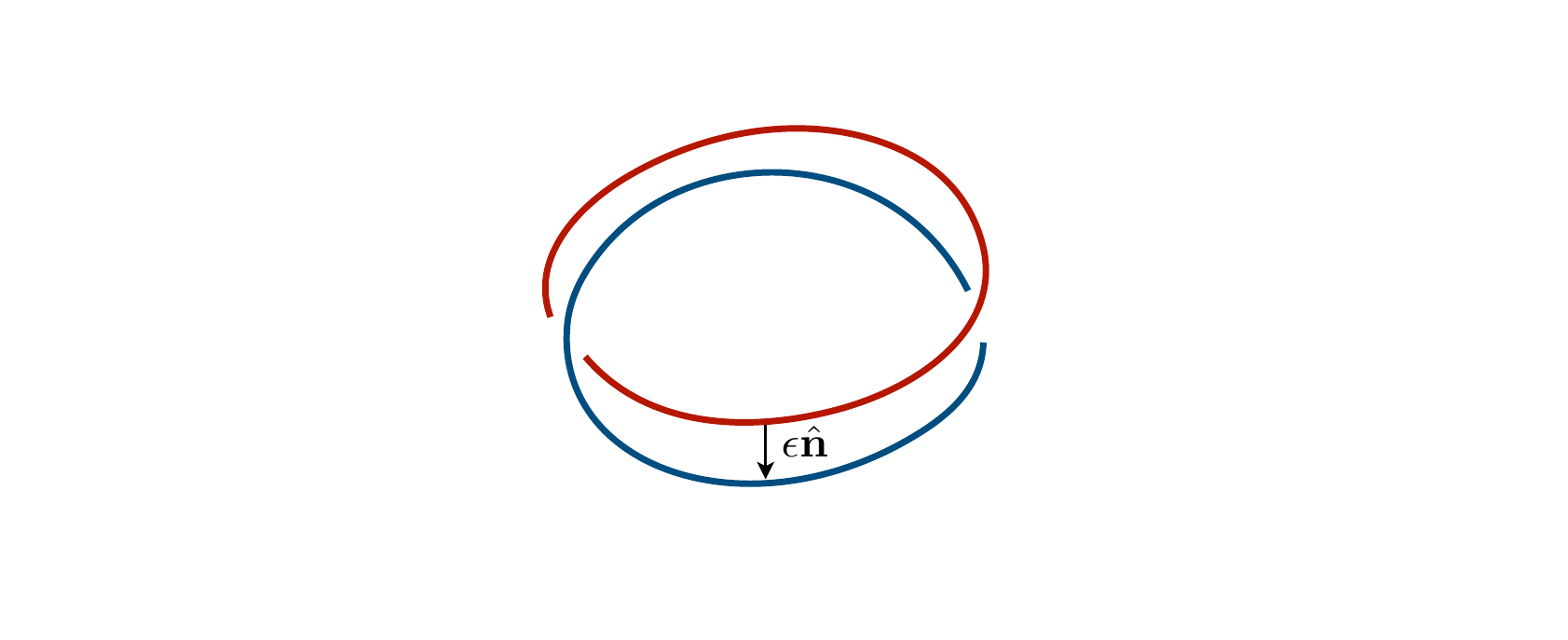}
\caption{Worldline transformed into a ribbon of self-linking number $SL=1$.}
\label{Fig:ribbon}
\end{figure}

As a first step, let us compute the expectation value of the Wilson loop when space-time is flat
\beq
\Phi_{flat}(C)=\lim_{e\rightarrow\infty}\,\exp\left\{\frac{e^2}{8\pi}\oint_C dx^a\oint_C dy^b\Big[i\,\epsilon_{abc}\frac{r^c}{\mu r^3}\big(1-(1+\mu r)e^{-\mu r}\big)-\delta_{ab}\frac{e^{-\mu r}}{r}\Big]\right\}.
\label{Eq. WL screened}
\eeq
By doing $r<<1/\mu$ we see that the first term goes as $\mu\epsilon_{abc}r^c/r$, so that the only singularity lives entirely in the screened Coulomb term $e^{-\mu r}/r$. Taking the limit $e\rightarrow\infty$ we get
\beq
\Phi_{flat}(C)=\exp\left[\frac{i}{4}\oint_C dx^a \oint_C dy^b \epsilon_{abc}\frac{r^c}{r^3}\right]=e^{i\pi W_{flat}}.
\label{Eq. writhe}
\eeq
The integral (\ref{Eq. writhe}) is finite and equal to the writhe \textit{W} of the curve \textit{C} \cite{F71}. The writhe depends on the shape of the curve, hence the Wilson loop depends on the metric of space-time. This might seem in contradiction with Witten's work \cite{W89}, where he regularized the integral \eqref{Eq. writhe} by means of a `point splitting' procedure which yields the result
\beq
\exp\left[\frac{i}{4}\oint_C dx^a\oint_C dy^b\epsilon_{abc}\frac{(r-\epsilon\hat{n})^c}{|\bm r-\epsilon\hat{\bm n}|^3}\right]=e^{i\pi SL},
\label{Eq. point splitting}
\eeq
where the unit normal vector $\hat{\bm n}$ defines a framing for the curve \textit{C}. This can be viewed as if the curve is transformed into a ribbon (see figure \ref{Fig:ribbon}). The integral of Eq.\eqref{Eq. point splitting} gives a topological invariant called self-linking number (SL), which is an integer and is basically the Gauss's linking number between the two edges of the ribbon. It depends on the choice of frame $\hat{n}$, and is related to the writhe \textit{W} by the formula $SL=W+T$, where \textit{T} is the twist of the ribbon \cite{W69}.

So we have two different regularizations of the Wilson loop: Eq. \eqref{Eq. writhe}, obtained as the $e\rightarrow\infty$ limit of Eq. (\ref{Eq. WL screened}), which is metric dependent but frame independent, and Eq.\eqref{Eq. point splitting}, obtained from the pure CS theory by doing point splitting, which is topological but frame dependent. This is a manifestation of the framing anomaly of the CS theory \cite{W89,BW91}, which reflects the impossibility of finding a regularization which is topological and frame independent at the same time. The reason definitions of Eqs. \eqref{Eq. writhe} and \eqref{Eq. point splitting} give different results lies in the fact that the limit $\epsilon\rightarrow0$ is not smooth. 
The ambiguity in this limit is lost in Eq. (\ref{Eq. WL screened}), where the $F^2$ term smoothens the singularity of the CS propagator, and the metric dependent result of Eq. \eqref{Eq. writhe} is obtained \cite{H89}. This is the correct result in our case, since we are considering the pure CS theory as the $e\rightarrow\infty$ limit of a Maxwell-CS theory. The $F^2/4e^2$ term in the gauge field Lagrangian plays the role of  a UV regulator which is forcing us to quantize the theory in a non topological, but frame invariant way.

Next let us introduce a moving frame $(\bm{e}^1,\bm{e}^i)$ along the worldline \textit{C} (with $i=2,3$) given in parametric form by $X_a(s)$, and
\beq
|\bm e^1|=|\bm e^i|=1;\quad e^1_a=\frac{\dot{X}_a}{|\dot{\bm X}|}=\epsilon_{abc}\,e^{2}_b\,e^{3}_c.
\eeq
The twist \textit{T} can be defined as \cite{P88}
\beq
T_{flat}=\frac{1}{2\pi}\int_0^t ds\,\epsilon_{abc}\,e^1_a\,e^{2}_b\,\dot{e}^{2}_c.
\label{eq twist}
\eeq
Following Refs. \cite{P88,P88LH,CL89,TJ89,KK91} we can write \textit{T}, and therefore \textit{W} through the formula $SL=W+T$, as a Wess-Zumino-Witten (``Berry phase'') term \cite{WZ71,WW-Z83} (see appendix \ref{sec:twist-as-berry})
\beq
W_{flat}=\frac{\Omega_{flat}}{2\pi}+SL+\nu.
\label{eq writhe 1}
\eeq
where the integer $\nu\in\mathbb{Z}$ is the self-winding number of the loop and the Wess-Zumino-Witten action, $\Omega_{flat}$, is
\beq
\Omega_{flat}=\int_0^1d\rho\int_0^t ds\, \epsilon_{abc}\,e^1_a\,\partial_{\rho}e^1_b\,\partial_{s}e^1_c.
\eeq
Here $\rho\in[0,1]$ is an auxiliary coordinate, and $\bm{e}^1$ is extended to $\bm{e}^1(t,\rho)$ in such a way that $\bm{e}^1(t,0)=\bm{e}_0$ and $\bm{e}^1(t,1)=\bm{e}^1(t)$, with $\bm{e}_0$ some constant vector. It turns out that $\Omega_{flat}$ is just the signed area enclosed by the path \textit{C} on the sphere $S_2$, defined by $\bm{e}^1: C\rightarrow S_2$. Finally, it can be seen that $W_{flat}-\Omega_{flat}/2\pi$ is an odd integer \cite{GH89,IIM90} , so that
\beq
W_{flat}=\frac{\Omega_{flat}}{2\pi}+2n+1, \quad n\in\mathbb{Z},
\eeq
and the Wilson loop reads
\beq
\Phi_{flat}(C)=-e^{i\Omega_{flat}/2}.
\eeq

As a final note, it is instructive to check the frame invariance of the Wilson loop. By doing the clockwise rotation
\beq
\tilde{\bm e}^2=\cos\varphi\,\bm e^2+\sin\varphi\,\bm e^3,\quad\quad \tilde{\bm e}^3=-\sin\varphi\,\bm e^2+\cos\varphi\,\bm e^3,
\eeq
where periodicity demands $\varphi(t)=\varphi(0)+2\pi m$, with $m\in\mathbb{Z}$, the winding number $\nu$ in Eq. (\ref{eq writhe 1}) is shifted by $\tilde{\nu}=\nu+m$ (see appendix \ref{sec:twist-as-berry}). At the same time, such a rotation of the frame also shifts the self-linking number: $\tilde{SL}=SL-m$. Therefore the writhe, and by extension the Wilson loop, are frame invariant.

\subsection{The Wilson loop in curved space-time}

Extending the Wilson loop, in the limit $e\rightarrow\infty$, to curved space-times is quite straight forward. The most obvious way to do it is to start with the definition in terms of the twist of the curve
\beq
\Phi(C)=e^{i\pi W}=e^{i\pi SL}e^{-i\pi T}.
\eeq
The extension of the twist to curved spaces is then performed by the substitution $\partial_s\rightarrow \dot{X}^\gamma\nabla_\gamma$, with $\nabla_\gamma\xi^\nu=\partial_\gamma\xi^\nu+\Gamma^\nu_{\beta\gamma}\xi^\beta$, where $\Gamma^\nu_{\beta \gamma}$ is the Christoffel symbol. We get
\beq
T=\frac{1}{2\pi}\int_0^t ds\,\epsilon_{\mu\rho\nu}\,e^{1\mu}\,e^{2\rho}\,\dot{X}^\gamma\nabla_\gamma e^{2\nu}.
\eeq
Using the vielbein postulate
\beq
\nabla_\gamma E^\nu_a=\partial_\gamma E^\nu_a+\Gamma^\nu_{\beta\gamma}E^\beta_a-\tensor{\omega}{_\gamma_a^b}E^\nu_b=0,
\eeq
with $E^\nu_a$ the vielbein and $\tensor{\omega}{_\gamma_a^b}$ the spin connection, we can write
\beq
\dot{X}^{\gamma}\nabla_\gamma e^{2\nu}=\dot{X}^{\gamma}\nabla_\gamma(E^\nu_ae^{2a})=E^\nu_a\dot{e}^{2a}+e^{2a}\dot{X}^{\gamma}(\partial_\gamma E^\nu_a+\Gamma^\nu_{\beta\gamma}E^\beta_a)=E^\nu_a\dot{e}^{2a}+e^{2a}\dot{X}^{\gamma}\tensor{\omega}{_\gamma_a^b}E^\nu_b.
\eeq
The twist then reads
\beq
T=T_{flat}+\frac{1}{2\pi}\int_0^t ds\,e^2_a\,e^3_b\,\dot{X}^\gamma\,\tensor{\omega}{_\gamma^a^b},
\eeq
where we have done $\epsilon_{\mu\rho\nu}\,e^{1\mu}\,e^{2\rho}=e^3_\nu$, and $T_{flat}$ is the twist in the non-coordinate basis, denoted by Latin indexes. Using the fact that the spin connection is antisymmetric under $a\leftrightarrow b$, we can write $e^2_a\,e^3_b=e^2_{[a}\,e^3_{b]}=\epsilon_{abc}\,e^{1c}/2$, and arrive to
\beq
T=T_{flat}+\frac{1}{4\pi}\int_0^t ds\,\dot{X}^\mu\,e^{1a}\,\epsilon_{abc}\,\tensor{\omega}{_\mu^b^c}.
\eeq
Finally, the expression for the Wilson loop in curved space-time is
\beq
\Phi(C)=-e^{i\Omega/2}
\eeq
\beq
\Omega=\Omega_{flat}-\frac{1}{2}\int_0^t ds\,\dot{X}^{\mu}e^{1a}\epsilon_{abc}\,\tensor{\omega}{_\mu^b^c}.
\eeq
So the Wilson loop divides into two contributions: the Wess-Zumino-Witten term, $\Omega_{flat}$, which is topological and does not depend on the metric, and the spin connection term, which encodes all the information of the background curvature. An alternative derivation in terms of path integrals of one dimensional fermions is given in appendix \ref{sec:spin-as-pathintegral}.

\subsection{Framing anomaly of the free CS theory}
\label{sec:framing-anomaly}

Now that we have the Wilson loop, we need to compute the vacuum contribution of the gauge field sector, $Z_{GF}$, which in the limit $e\rightarrow\infty$ is just the partition function for the pure CS theory, $Z_{CS}$\footnote{Actually, the term $F^2/4e$ in $Z_{GF}$ plays the role of a regulator for the CS theory. Strictly speaking, there will appear metric dependent terms in $Z_{GF}$ that go as positive powers of $e$ \cite{BW91}, which diverge in the limit $e\rightarrow\infty$ and need to be taken care by counterterms.}. At the classical level, the CS theory is well known to be topological, but this is no longer true when one considers the quantum theory. In the path integral formulation, this means that the integration measure in the partition function is metric dependent. This dependence arises from the need to fix the gauge, which can be done by including gauge fixing terms which do depend on the metric. The calculation of $Z_{CS}$ in a curved background was first done by Witten \cite{W89}, so here we  just recall the result \cite{W89,BW91,GF15}
\beq
Z_{CS}=\tau^{1/2}\,e^{-i\pi\eta_{g}/2}.
\label{eq eta conclusions}
\eeq
$\tau^{1/2}$ is a topological invariant, the square root of the Ray-Singer analytic torsion \cite{RS71}, which is just a constant multiplicative term. Here $\eta_{g}$ is the eta-invariant \cite{APS75} of the purely gravitational operator. It is a regularized version of the difference between the number of positive and negative eigenvalues of the Dirac operator. In the special case that our three-dimensional manifold is a boundary of some four-dimensional manifold, then the Atiyah-Patodi-Singer  index theorem \cite{APS75} tells us that \cite{W16}
\beq
\frac{\eta_g}{2}=\frac{CS_g}{\pi}\,\,\mathrm{mod}\,\mathbb{Z}
\eeq
Here $CS_g$ is the gravitational CS term \cite{DJT82,W88}, which in terms of the spin connection is given by
\beq
CS_{g}=\frac{1}{96\pi}\mathrm{tr}\int d^3x\,\epsilon^{\mu\rho\nu}\big(\omega_\mu\partial_\rho\omega_\nu+\frac{1}{3}\omega_\mu[\omega_\rho,\omega_\nu]\big)
\eeq
For a general three-dimensional manifold, a topological invariant can be constructed by subtracting $CS_g$ to the eta-invariant: $\eta_g/2-CS_g/\pi$. The subtlety is that $CS_g$ depends on the choice of frame, so the quantity $\eta_g/2-CS_g/\pi$ can be defined as a topological invariant of a framed manifold \cite{W89,BW91}. This is what gives rise to the framing anomaly, because one could think of regulating the theory by adding a counterterm $-CS_g$ to make $Z_{CS}$ topological, but this would come with the price of loosing frame invariance. 

Finally, from the previous discussion we can set $\eta_g/2=CS_g$, which is always true up to multiplicative constant terms unimportant for our discussion. We can then write
\beq
Z_{CS}=e^{-iCS_{g}}.
\label{partition function framing anomaly}
\eeq

\subsection{Effective action}
Putting the previous pieces together, we obtain to the following effective action, in the limit $e\rightarrow\infty$
\begin{align}
\Gamma_{scalar+flux}&=iCS_g\nonumber\\
+\int_{\epsilon}^\infty &dt\,t^{-1}\,\int DX^\mu\,e^{-\int_0^t ds\big(\frac{1}{4}\dot{X}^2-iA_\mu\dot{X}^\mu+\frac{i}{4}e^{1a}\epsilon_{abc}\,\tensor{\omega}{_\mu^b^c}\dot{X}^{\mu}+\frac{R}{4}+m^2+2i\sqrt{\lambda}\,\sigma\big)}\, e^{i\Omega_{flat}/2}\,. 
\label{eq:duality-boson}
\end{align}
We have added the contribution of the framing anomaly $iCS_g$, so the notation now reads 
\begin{equation}
Z^{e\rightarrow\infty}_{scalar+flux}=\int D\sigma\,e^{-\int d^3x\,\sigma^2}\,e^{-\Gamma_{scalar+flux}}.
\end{equation}

\section{Partition function for a Dirac fermion}
\label{sec:effactionfermions}
We will now consider a massive Dirac fermion in 2+1 dimensions, and write its effective action as a worldline path integral. The goal is to compare the result with the effective action for the relativistic composite boson. In order to do this, we can write the action of a Dirac fermion as
\beq
S_f=\int d^3x\,\sqrt{g}\,\bar{\Psi}\big(i\slashed D(A,\omega)-m\big)\Psi,
\label{eq action dirac fermion Minkowski}
\eeq
with $\slashed D=\gamma^\mu D_\mu$, $\gamma^\mu$ the Gamma matrices, and
\beq
D_\mu(\omega,A)=\partial_\mu+iA_\mu+\frac{1}{8}\tensor{\omega}{_\mu^a^b}[\gamma_a,\gamma_b].
\eeq
The partition function reads
\beq
Z_f=\int D\bar{\Psi}D\Psi\,\,e^{iS_f}=\det \Big(\slashed D(A,\omega)+im\Big).
\eeq
Wick rotating to three-dimensional Euclidean space-time it becomes
\beq
Z_f=e^{-\Gamma_f}=\det \Big(i\slashed D(A,\omega)+im\Big),
\eeq
where $\slashed D=\sigma^\mu D_\mu$, with $\sigma^\mu$ the Pauli matrices, and the metric is now given in euclidean signature. The covariant derivative reads
\beq
D_\mu(\omega,A)=\partial_\mu+iA_\mu+\frac{1}{8}\tensor{\omega}{_\mu^a^b}[\sigma_a,\sigma_b].
\eeq
We can separate the real and imaginary parts of the effective action as follows
\beq
\Gamma_f=\mathcal{R}(\Gamma_f)+i\mathcal{I}(\Gamma_f)=-\ln\big|\det\big(i\slashed D(A,\omega)+im\big)\big|-i\,\mathrm{arg}\Big[\det(i\slashed D(A,\omega)+im)\Big],
\eeq
where $arg$ denotes the argument. By using the identities $|\det\hat{\mathcal{O}}|^2=\det(\hat{\mathcal{O}}^\dagger\hat{\mathcal{O}})$ and $\ln\det=\mathrm{tr}\ln$, we get
\beq
\mathcal{R}(\Gamma_f)=-\frac{1}{2}\mathrm{tr}\ln\Big(-\slashed D^2(A,\omega)+m^2\Big)
\eeq
\beq
\mathcal{I}(\Gamma_f)=-\mathrm{arg}\Big[\det(i\slashed D(A,\omega)+im)\Big].
\eeq
The imaginary part of the effective action breaks time-reversal and parity symmetry. This comes from the fact that, in an orientable space-time $X$ of euclidean signature, the partition function  complex conjugates under a time-reversal transformation, which can be implemented by reversing the orientation of $X$ \cite{W16}.

The imaginary, time-reversal breaking part of the effective action of a Dirac fermion is well known to acquire a term proportional to the combination $-CS_A/2-CS_g$, the proportionality constant being the sign of the fermion mass\footnote{Actually, the term generated by integrating out fermions is $-\mathrm{sign}(m)\pi\eta/2$. From now on, we approximate the $\eta$-invariant by $CS_A/\pi+2CS_g/\pi$.} \cite{NS83,R84,GPM85,GV86,V86,BPR86,SO89}. In addition, there is an extra contribution $\mp CS_A/2\mp CS_g$ due to the parity anomaly \cite{R84,GPM85,GV86,V86,GRS96,KV18}, the sign depending on the choice of regularization, e.g. if one regularizes using a Pauli-Villars regulator, the sign corresponds to the sign of the regulator mass. Putting the two contributions together one gets\footnote{This expression is valid only in the deep IR for fixed values of the mass $m$. For finite values of the mass there is, in principle, an infinite series of parity-odd operators whose coupling scales as $m^{-r}$, with $r$ an odd positive integer. Here we will use this expression as an approximate result. For an alternative view see Ref. \cite{GRS96}.}
\beq
\mathcal{I}(\Gamma_f)=-\big(\mathrm{sign}(m)\pm1\big)\Big(\frac{CS_A}{2}+CS_g\Big)
\eeq

Regarding the real part of the effective action, we make use of the Schr\"odinger-Lichnerowicz formula for the squared Dirac operator\footnote{We are assuming vanishing torsion ($\Gamma^\mu_{[\rho\nu]}=0$), otherwise additional torsion terms would appear in Eq. (\ref{eq:S-Lformula}).}
\beq
-\slashed D^2(A,\omega)=-\frac{1}{\sqrt{g}}D_\mu(A,\omega)\sqrt{g}\,g^{\mu\nu}D_\nu(A,\omega)+\frac{R}{4}-\frac{i}{4}F_{\mu\nu}[\sigma^\mu,\sigma^\nu].
\label{eq:S-Lformula}
\eeq
Then, using results of appendix \ref{sec:worldline}, we can write it as a worldline path integral
$$
-2\mathcal{R}(\Gamma_f)=\mathrm{tr}\ln\big(-\slashed D^2(A,\omega)+m^2\big)
$$
\beq
=-\int_{\epsilon}^\infty dt\,t^{-1}\,\int DX^\mu\,\tilde{\mathrm{tr}}\,\mathcal{P}\,e^{-\int_0^tds\big(\frac{1}{4}\dot{X}^2-iA_\mu\dot{X}^\mu+F^*_\mu(A)\sigma^\mu+\frac{i}{4}\tensor{\omega}{_\mu^a^b}\epsilon_{abc}\sigma^c\dot{X}^\mu+\frac{R}{4}+m^2\big)},
\label{Eq. real effective action fermion}
\eeq
where the tilde in $\tilde{\mathrm{tr}}$ means the trace is over the spin indices, $\mathcal{P}$ denotes path ordering, and $F^*_\mu(A)$ is the Hodge dual of the background field strength $F^*_\mu(A)=\epsilon_{\mu\rho\nu}F^{\rho\nu}(A)/2$. 

To compute the trace, we follow Ref.\cite{GHK90} and parametrize the states by picking the ``time'' direction, $x_3$, and ``boost'' eigenstates of $\sigma^3$ to eigenstates of $e^1_\mu\sigma^\mu$: $e^1_\mu\sigma^\mu\,|\bm e^1,+\rangle=|\bm e^1,+\rangle$, $e^1_\mu\sigma^\mu\,|\bm e^1,-\rangle=-|\bm e^1,-\rangle$. The trace of a spin operator then reads $\tilde{\mathrm{tr}}\,\hat{\mathcal{O}}=\langle \bm e^1,+|\hat{\mathcal{O}}|\bm e^1,+\rangle+\langle \bm e^1,-|\hat{\mathcal{O}}|\bm e^1,-\rangle$. For an operator of the form $\hat{\mathcal{O}}=e^{i\int_0^tds\frac{1}{2}B_\mu\sigma^\mu}$, the quantity $\langle \bm e^1,\pm|\hat{\mathcal{O}}|\bm e^1,\pm\rangle$ has been evaluated in the context of coherent state path integrals for non relativistic spinning particles \cite{FS88,Wieg88,S89,O89,FMS92}
\beq
\langle \bm e^1,\pm|e^{i\int_0^tds\frac{1}{2}B_\mu\sigma^\mu}|\bm e^1,\pm\rangle=e^{\pm i\int_0^tds\frac{1}{2}B_\mu e^{1\mu}}e^{\pm\frac{i}{2}\Omega_{flat}}.
\eeq
Using this result, we get for $\mathcal{R}(\Gamma_f)$
\begin{align}
\mathcal{R}(\Gamma_f)&=\int_{\epsilon}^\infty dt\,t^{-1}\,\int DX^\mu\,e^{-\int_0^tds\big(\frac{1}{4}\dot{X}^2-iA_\mu\dot{X}^\mu+\frac{R}{4}+m^2\big)}\nonumber\\
&\times\cos\Big[\int_0^tds\,\Big(iF^*_\mu(A)e^{1\mu}-\frac{1}{4}e^{1a}\epsilon_{abc}\tensor{\omega}{_\mu^b^c}\dot{X}^\mu\Big)+\frac{\Omega_{flat}}{2}\Big].
\label{eq:duality-fermions}
\end{align}
The negative of  the cosine factor in Eq. \eqref{eq:duality-fermions} is Polyakov's spin factor \cite{P88LH,GHK90}, it is what gives the path integral the fermionic character. The spin factor can be written, alternatively, as a path integral over anticommuting fields (see appendix \ref{sec:spin-as-pathintegral}).

\section{Boson-fermion duality}
\label{sec:dualitytimereversal}

\begin{figure}
\centering
\includegraphics[scale=0.28]{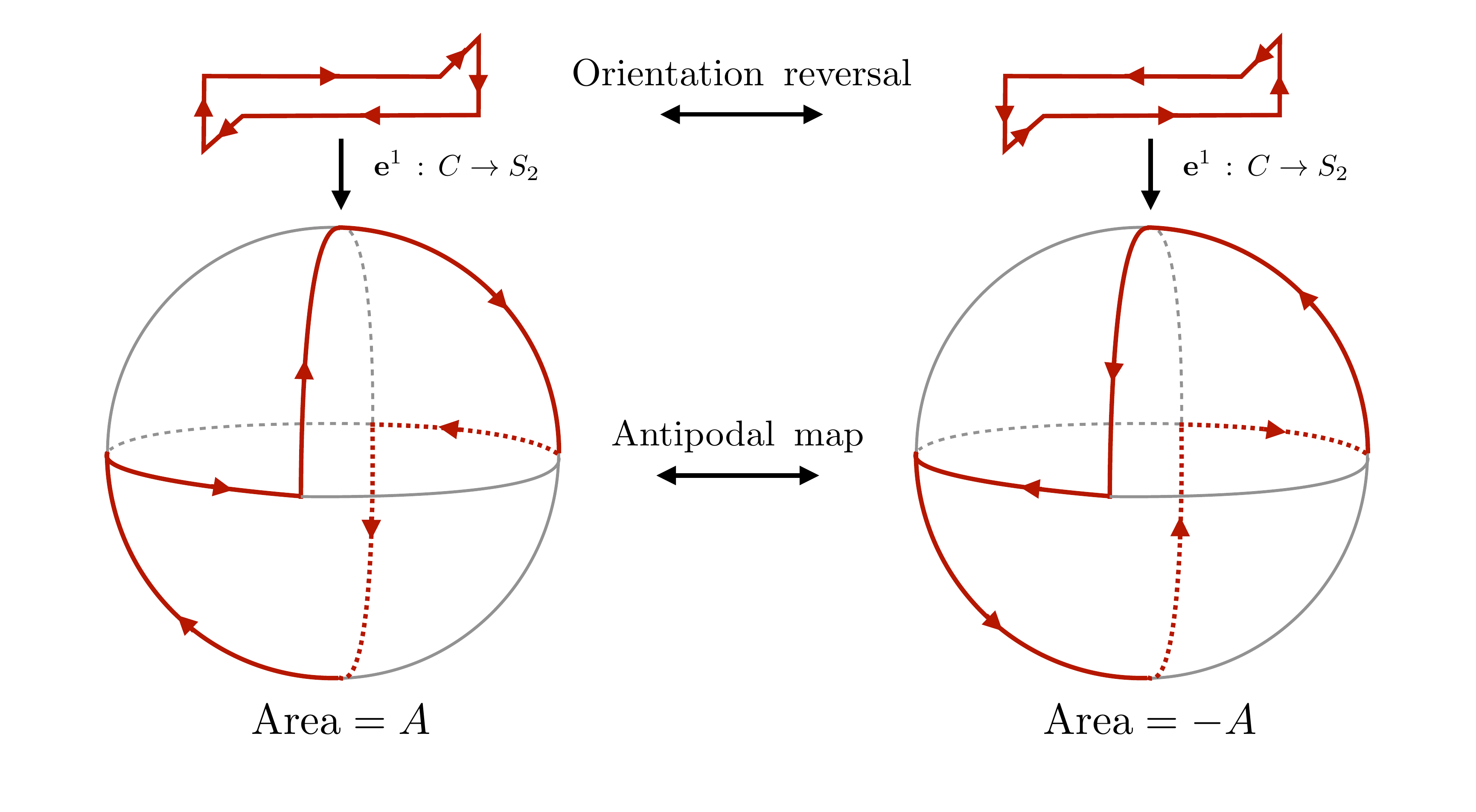}
\caption{Schematics of two paths, related by reversal of orientation, and their mappings on the sphere. The implementation of orientation reversal on the sphere is realized by an antipodal map: $\mathcal{A}(\bm e^1)=-\bm e^1$. The enclosed signed areas of the paths are $2\pi$ and $-2\pi$.}
\label{Fig:antipodal}
\end{figure}

Let us look more carefully at the effective action for the relativistic composite boson, Eq. \eqref{eq:duality-boson}, and do the following reasoning. Remember that $\Omega_{flat}$ is the signed area enclosed by a given worldline $C$ on the sphere $S_2$, defined by $\bm e^1:C\rightarrow S_2$. Now suppose that we want to reverse the orientation of $C$. We can do it by implementing an antipodal map on the sphere, this is, under orientation reversal each point of a given path $C$ on $S_2$ is mapped to its antipodal point. This is easy to see by noting that reversing orientation is basically flipping the sign of the unit tangent vector at each point on $C$. Thus, any path on $S_2$ is mapped to its antipodal path, see figure \ref{Fig:antipodal}. If the signed area enclosed by a given path on $S_2$ is $A$, then the area of the antipodal path is $-A$. This means that under orientation reversal, $\Omega_{flat}\rightarrow-\Omega_{flat}$ and $\exp(i\Omega_{flat}/2)$ is complex conjugated.
Having this in mind, we perform the following trivial manipulation
\begin{align}
e^{\frac{i}{2}\Omega_{flat}}\,e^{i\int_0^tds\,\big(A_\mu\dot{X}^\mu-\frac{1}{4}e^{1a}\epsilon_{abc}\,\tensor{\omega}{_\mu^b^c}\dot{X}^{\mu}\big)}=&\,e^{i\int_0^tds\,A_\mu\dot{X}^\mu}\cos\Big(\frac{\Omega_{flat}}{2}-\int_0^tds\,\frac{1}{4}e^{1a}\epsilon_{abc}\,\tensor{\omega}{_\mu^b^c}\dot{X}^{\mu}\Big)
\nonumber\\
+\Big(i\cos\int_0^tds\,A_\mu\dot{X}^\mu-\sin &\int_0^tds\,A_\mu\dot{X}^\mu\Big) \sin\frac{\Omega_{flat}}{2}\,\cos\Big(\frac{1}{4}\int_0^tds\,e^{1a}\epsilon_{abc}\,\tensor{\omega}{_\mu^b^c}\dot{X}^{\mu}\Big)
\nonumber\\
-\Big(i\cos\int_0^tds\,A_\mu\dot{X}^\mu-\sin &\int_0^tds\,A_\mu\dot{X}^\mu\Big)\cos\frac{\Omega_{flat}}{2}\,\sin\Big(\frac{1}{4}\int_0^tds\,e^{1a}\epsilon_{abc}\,\tensor{\omega}{_\mu^b^c}\dot{X}^{\mu}\Big).
\label{eq:sinesandcosines}
\end{align}
Contrary to $\Omega_{flat}$ and $\int_0^tds\,A_\mu\dot{X}^\mu$, the spin connection term remains invariant under reversal of orientation. This is of no surprise, since reversing orientation of the worldlines can be seen as a charge conjugation operation. Applying Eq. \eqref{eq:sinesandcosines} to the effective action \eqref{eq:duality-boson}, and after integrating over all worldline configurations, which include both forward and backward orientations for each path, odd terms under orientation reversal vanish and the r.h.s. of Eq.\eqref{eq:sinesandcosines} becomes
\begin{align}
&e^{i\int_0^tds\,A_\mu\dot{X}^\mu}\cos\Big(\frac{\Omega_{flat}}{2}-\int_0^tds\,\frac{1}{4}e^{1a}\epsilon_{abc}\,\tensor{\omega}{_\mu^b^c}\dot{X}^{\mu}\Big)-\sin\int_0^tds\,A_\mu\dot{X}^\mu\,\sin\frac{\Omega_{flat}}{2}
\nonumber\\
&\times\cos\Big(\frac{1}{4}\int_0^tds\,e^{1a}\epsilon_{abc}\,\tensor{\omega}{_\mu^b^c}\dot{X}^{\mu}\Big)-i\cos\int_0^tds\,A_\mu\dot{X}^\mu\,\cos\frac{\Omega_{flat}}{2}\,\sin\Big(\frac{1}{4}\int_0^tds\,e^{1a}\epsilon_{abc}\,\tensor{\omega}{_\mu^b^c}\dot{X}^{\mu}\Big).
\label{eq:sinesandcosines2}
\end{align}
The effective action must preserve charge conjugation symmetry, which means that odd terms in $A_\mu$ should vanish (by Furry's theorem). Then the second term in Eq. \eqref{eq:sinesandcosines2} vanishes. The third term makes up the imaginary part of the bosonic effective action. It contains only odd powers of the spin connection, and because there is no locally Lorentz invariant term that can be constructed with odd powers of $\omega_\mu$ alone\footnote{The Lagrangian (\ref{eq action composite boson}) for the relativistic composite boson has local Lorentz invariance, so the effective action should share this feature. Local Lorentz invariance restricts terms in the effective action to be a functional of the curvature $\tensor{R}{^a_b_\mu_\nu}=\partial_\mu\tensor{\omega}{_\nu^a_b}-\partial_\nu\tensor{\omega}{_\mu^a_b}+\tensor{\omega}{_\mu^a_c}\,\tensor{\omega}{_\nu^c_b}-\tensor{\omega}{_\nu^a_c}\,\tensor{\omega}{_\mu^c_b}$. The exception is the CS term, which is a functional of both the curvature and spin connection: $CS_g\propto\epsilon^{\mu\rho\nu}\,\tensor{\omega}{_\mu^a_b}\,\tensor{R}{^b_a_\rho_\nu}$. The CS term is locally Lorentz invariant only up to a total derivative.}, it should vanish as well. We are thus left with
\beq
e^{\frac{i}{2}\Omega_{flat}}\,e^{i\int_0^tds\,\big(A_\mu\dot{X}^\mu-\frac{1}{4}e^{1a}\epsilon_{abc}\,\tensor{\omega}{_\mu^b^c}\dot{X}^{\mu}\big)}=e^{i\int_0^tds\,A_\mu\dot{X}^\mu}\cos\Big(\frac{\Omega_{flat}}{2}-\int_0^tds\,\frac{1}{4}e^{1a}\epsilon_{abc}\,\tensor{\omega}{_\mu^b^c}\dot{X}^{\mu}\Big).
\label{eq:sinesandcosines3}
\eeq

\subsection{Duality as a mapping between phases}

In order to be able to compare bosonic and fermionic effective actions we note that the $|\phi|^4$ interaction is relevant, which means that its strength grows as we run to the IR. For massive excitations, we can assume our energy scale to be sufficiently far away from the IR, such that treating the interactions perturbatively seems reasonable. Doing so and inserting Eq. \eqref{eq:sinesandcosines3} in \eqref{eq:duality-boson} we obtain, for the effective action for the relativistic composite boson\footnote{Please, note the slight change in notation compared to Eq. \eqref{eq:duality-boson}. Now $\Gamma_{scalar+flux}$ contains the short-range interacting terms resulting by integrating out $\sigma$.}
$$
Z_{scalar+flux}=e^{-\Gamma_{scalar+flux}}
$$
$$
\Gamma_{scalar+flux}=\int_{\epsilon}^\infty dt\,t^{-1}\,\int DX^\mu\,e^{-\int_0^tds\big(\frac{1}{4}\dot{X}^2-iA_\mu\dot{X}^\mu+\frac{R}{4}+m^2\big)}
$$
\beq
\times\cos\Big[-\frac{1}{4}\int_0^tds\,e^{1a}\epsilon_{abc}\tensor{\omega}{_\mu^b^c}\dot{X}^\mu+\frac{\Omega_{flat}}{2}\Big]+iCS_g+\mathcal{O}(\lambda),
\label{eq effective action boson final}
\eeq
where $\mathcal{O}(\lambda)$ includes (short-range) interacting terms at all non-trivial orders in $\lambda$. By comparison with the real part of the fermion effective action, Eq. \eqref{eq:duality-fermions}, we see that the Pauli term in the spin factor (inside the cosine), proportional to $F^*_\mu(A)$, is missing in Eq. \eqref{eq effective action boson final}. Such a term is a consequence of the spin angular momentum of the Dirac spinor, and contributes to the even part of the vacuum polarization in the fermionic side (and in general to the \textit{N}-photon amplitudes). While it can not be reproduced in the bosonic side, $F^*_\mu(A)$ is an irrelevant operator, which means that its value decreases towards the IR, vanishing at the IR fixed point. We can write
\beq
\mathcal{R}(\Gamma_{scalar+flux})\approx\mathcal{R}(\Gamma_f)+\mathcal{O}(\lambda),
\eeq
where $\approx$ means that the equality is only valid up to irrelevant terms of the order $\mathcal{O}\big(F^*_\mu(A)\big)$.

\subsubsection{Duality as a mapping between topological phases}

Turning our attention to the map between imaginary parts of effective actions, let us first pick a specific regularization for the parity anomaly of the fermion, say
\beq
\mathcal{I}(\Gamma_f)=-\big(\mathrm{sign}(m)-1\big)\Big(\frac{CS_A}{2}+CS_g\Big).
\label{eq:imaginary-fermion-fixed-reg}
\eeq
We see that, if the mass is negative, Eq. \eqref{eq:imaginary-fermion-fixed-reg} does not match the imaginary part of the $\Gamma_{scalar+flux}\,$. On the one hand, the electromagnetic CS term is missing in $\Gamma_{scalar+flux}\,$, so to obtain an equality between imaginary parts of effective actions we need to add a counterterm of the form $iCS_A$ to the bosonic side. On the other hand, there is a mismatch between the gravitational contributions. The framing anomaly as in Eq. \eqref{partition function framing anomaly} gives a contribution $iCS_g$ to $\Gamma_{scalar+flux}\,$, whereas the fermionic gravitational contribution in \eqref{eq:imaginary-fermion-fixed-reg} is twice as big, $i2CS_g$. We need to add a gravitational CS term $iCS_g$, which we can think of as an alternative regularization of the framing anomaly, so that the partition function of the free CS theory takes the form $Z_{CS}=e^{-i2CS_g}$. This choice of regularization for the framing anomaly has been recently used in this context \cite{SW16}, in order to obtain the correct boson-fermion mapping. Adding the above two terms we get
\beq
\Gamma_{scalar+flux}+iCS_A+iCS_g\approx\Gamma_f+\mathcal{O}(\lambda)
\label{eq effective action boson real 2}
\eeq
In Lagrangian form this translates to\footnote{To avoid further clarifications on whether a given expression is given in Minkowski or Euclidean signature, we note that whenever we write classical actions or Lagrangians we assume Minkowski signature for the metric, whereas effective actions are given in Euclidean signature.}
$$
\big|D_\mu(a)\phi\big|^2-(\frac{R}{4}+m^2)|\phi|^2-\lambda |\phi|^4-\frac{1}{4e}F^2(a)+CS_a-CS_g+\frac{1}{2\pi}\epsilon^{\mu\rho\nu}a_\mu\partial_\rho A_\nu
$$
$$
\xleftrightarrow{e\rightarrow\infty}
$$
\beq
\bar{\Psi}\big(i\slashed D(\omega,A)-m\big)\Psi+\,.\,.\,.
\label{eq:duality-general}
\eeq
where we have done the shift $a_\mu\rightarrow a_\mu-A_\mu$, and where $m$ is negative: $m=-|m|$. By $\,.\,.\,.$ we denote short-range fermionic interactions, that account for the short-range interactions, $\mathcal{O}(\lambda)$, in the bosonic side. It's important to note that the correspondence in Eq. \eqref{eq:duality-general} is only valid up to induced irrelevant terms of the order $\mathcal{O}\big(F^*_\mu(A)\big)$ in the fermionic effective action.

Eq. \eqref{eq:duality-general} is a mapping between theories in a topological  phase (usually referred to as the Chern insulator in condensed matter physics). The bosonic side flows, when the mass is taken to vanish, towards the IR Wilson-Fisher-CS fixed point, which is a strongly short-range interacting theory of massless bosons coupled to a CS field. So, even if the bare value of the quartic coupling is not tuned to the Wilson-Fisher fixed point, the RG flows in the IR to this fixed point. In other words, the fixed point theory and the bare theory differ in irrelevant operators that, at most, yield corrections to scaling.

Regarding the fermionic side, in the free Dirac theory the short-range four-fermion interactions (denoted by $.\,.\,.$) are irrelevant, and consequently vanish in the IR. Thus, in the IR we flow to the free massless Dirac fermion fixed point. The same happens with the irrelevant terms in $F^*_\mu(A)$ in the fermionic effective action that spoil the mapping \eqref{eq:duality-general}. Therefore, our results are consistent with the conjectured exact duality between a Wilson-Fisher-CS complex scalar and a free massless Dirac fermion at the IR fixed point \cite{KT16,SW16}.

It is important to emphasize that in a construction of both theories along the lines used in this work (and, in fact, in any local construction) the resulting theory is not automatically a CFT since, as we showed here, there are always irrelevant operators. Thus, the duality mappings are asymptotic statements of both theories deep in the IR. That this is the case is an assumption of the CFT constructions of the dualities.

\subsubsection{Duality as a mapping between trivial phases}

Now we can think of flipping the sign of the fermion mass, this time to be positive. In this scenario the imaginary part of the fermionic effective action vanishes, and the fermionic side is now in a topologically trivial phase. This situation is understood to correspond to the U(1) symmetry broken (Higgs) phase of the composite boson \cite{KT16,SW16}. This has very recently been proven, in the context of loop models, by Goldman and one of us \cite{GF18} by making use of the particle-vortex duality, and writing the loop model partition function in the broken phase as one describing the symmetric phase of dual variables $\tilde{\phi}$. We are going to use the same procedure here in order to obtain the boson-fermion duality in the trivial phase. First, we invoke the mentioned particle-vortex mapping
\beq
\big|D_\mu(B)\phi\big|^2+m^2|\phi|^2-\lambda |\phi|^4\longleftrightarrow \big|D_\mu(\tilde{b})\tilde{\phi}\big|^2-m'^2|\tilde{\phi}|^2-\lambda' |\tilde{\phi}|^4+\frac{1}{2\pi}\epsilon^{\mu\rho\nu}\tilde{b}_\mu\partial_\rho B_\nu.
\eeq
The map identifies the Higgs phase of $\phi$ variables (note the positive sign in front of $m^2$) with the symmetric phase of $\tilde{\phi}$ variables. We can add now $CS_B+\epsilon^{\mu\rho\nu}B_\mu\partial_\rho A_\nu/2\pi$ to both sides, and promote $B_\mu$ to a dynamical field $b_\mu$. After integrating out $b_\mu$ on the r.h.s. we obtain
$$
\big|D_\mu(b)\phi\big|^2+m^2|\phi|^2-\lambda |\phi|^4+CS_b+\frac{1}{2\pi}\epsilon^{\mu\rho\nu}b_\mu\partial_\rho A_\nu
$$
$$
\longleftrightarrow
$$
\beq
\big|D_\mu(\tilde{b})\tilde{\phi}\big|^2-m'^2|\tilde{\phi}|^2-\lambda'|\tilde{\phi}|^4-CS_{\tilde{b}}-CS_A-\frac{1}{2\pi}\epsilon^{\mu\rho\nu}\tilde{b}_\mu\partial_\rho A_\nu
\eeq
We can make use of this duality and write the composite boson, appearing in the bosonic side of duality \eqref{eq:duality-general}, in its Higgs phase in terms of the $\tilde{\phi}$ field in its symmetric phase. We get
\beq
\big|D_\mu(\tilde{a})\tilde{\phi}\big|^2-(\frac{R}{4}+m^2)|\tilde{\phi}|^2-\lambda |\tilde{\phi}|^4-\frac{1}{4e}F^2(\tilde{a})-CS_{\tilde{a}}-CS_A-CS_g-\frac{1}{2\pi}\epsilon^{\mu\rho\nu}\tilde{a}_\mu\partial_\rho A_\nu
\label{eq:composite_boson_dual}
\eeq
It is clear that the real part of the effective action arising from \eqref{eq:composite_boson_dual} is basically that of the $\phi$ fields in the symmetric phase, $\mathcal{R}(\Gamma_{scalar+flux})$, as given by Eq. \eqref{eq effective action boson final}. On the other hand, the sign flip in the CS term for the dynamical gauge field $\tilde{a}_\mu$ gives a framing anomaly with gravitational CS term of opposite sign, $CS_g$, canceling the background contribution $-CS_g$. A similar thing happens for $A_\mu$: doing the shift $\tilde{a}_\mu\rightarrow\tilde{a}_\mu+A_\mu$, the resulting CS term, $CS_A$, is canceled by the extra term $-CS_A$. This means that the effective action arising from Eq. \eqref{eq:composite_boson_dual} is real, and is therefore equal to the effective action of a Dirac fermion of positive mass (with the chosen specific regularization of the parity anomaly). We have then proven that the composite boson $\phi$ in its Higgs phase is dual to a Dirac fermion of positive mass.

\section{Concluding remarks and outlook}
\label{sec:conclusions}

We have revisited Polyakov's duality between a massive relativistic composite boson and a massive Dirac fermion, in the presence of background curvature and electromagnetic fields. The duality holds in curved space-time, as long as a non-minimal coupling to curvature is included in the bosonic side, but holds only up to an irrelevant operator when fermions and bosons are coupled to a background gauge field. This is rooted to the obtained spin factor for the Dirac fermion (see Eq. \eqref{eq:duality-fermions}), which has both electromagnetic and gravitational contributions. While the gravitational contribution is provided naturally, in the bosonic side, by the extension of the Wilson loop to curved backgrounds, the electromagnetic contribution (Pauli term) is a consequence of the spin angular momentum of the spinor and can not be reproduced in the bosonic side.

The addition of background fields to Polyakov's construction has to be treated with care, as now the parity and framing anomaly of fermions and CS fields come in to fore. The mapping between partition functions can only be defined after proper regularizations for the anomalies have been specified. For the parity anomaly of the fermion, gauge invariance must be imposed, and a regularization dependent time-reversal breaking CS term appears in the effective action. Further fixing the sign of the CS term completely removes the ambiguity on the regularization of the parity anomaly. In regards to the framing anomaly, we included a Maxwell term, $F^2/e^2$, for the CS gauge field. Its purpose is to act as a regulator that forces the CS theory to be quantized in a non-topological, but frame invariant, way. The duality is then defined after removing the regulator, i.e. at infinite coupling $e\rightarrow\infty$. A metric dependent regularization of the CS theory is necessary in two levels: first, it is needed to obtain the spin factor from the Wilson loop, responsible of transmuting a boson into a fermion, and second, it is necessary to obtain the gravitational CS term from the framing anomaly of the pure CS theory, such that the imaginary parts (in Euclidean signature) of the effective
actions at both sides of the duality match.

The inclusion of $|\phi|^4$ interactions in the bosonic side is of crucial importance to resolve another subtlety of Polyakov's mapping. The machinery behind the boson-fermion duality relies on a topological property of Wilson loops in the CS theory. Basically, closed worldline configurations are classified in terms of the Gauss's linking number, such that interactions between worldlines vanish for specific values of the statistical angle: $\delta=n\pi$, with $n\in\mathbb{Z}$ (in our case, $\delta=\pi$). The subtlety lies in the fact that, in the absence of interactions between bosons, worldlines could intersect, in which case the linking number would not be well defined. Other way of seeing this is that if intersections are not avoided, linked worldlines and knots could be unlinked and unknotted, and their topological properties would not be preserved. The $|\phi|^4$ term solves this issue, providing short-ranged interactions between worldlines that suppress intersecting worldline configurations. The short-range bosonic interactions map to short-range interactions in the fermionic side, with the crucial difference that, unlike the $|\phi|^4$ interactions, which are relevant, short-range fermionic interactions are irrelevant. Thus, when the mass is set to zero, the fermionic side flows to a non-interacting Dirac fermion at the IR fixed point, while the bosonic side flows to a (strongly interacting) Wilson-Fisher-CS complex scalar.

The massive fermionic and bosonic theories at both sides of the duality can be either in a trivial or a topological phase. If the fermion is in the topological phase, it maps to a composite boson in its $U(1)$ symmetric phase, whereas if it is in the trivial phase, the mapping is to a composite boson in its $U(1)$ symmetry broken (Higgs) phase. To prove this last case, instead of working directly in the Higgs phase we followed Ref. \cite{GF18} and made use of the particle-vortex duality, writing the Higgs phase of the $\phi$ field as the $U(1)$ symmetric phase of a dual field $\tilde{\phi}$. Once working in the dual picture, the proof of the duality is analogous to that for the $\phi$ field in its symmetric phase.

Our work opens the path for computing gravitational and geometric responses of relativistic composite bosons. Some responses in condensed matter systems, like the Hall viscosity, can acquire torsional contributions. In solids, torsion can arise naturally in the form of dislocations. Henceforth, one interesting extension of the present work is to allow for a background with torsion. Another future appealing direction is to extend the formalism presented here to prove dualities involving non-Abelian CS-matter theories. Specifically, a recent conjectured duality between a Majorana fermion and a $SO(N)$ vector boson coupled to a $SO(N)$ CS gauge field \cite{ABHS17,MVX17} could be of relevance in a condensed matter context.



\section*{Acknowledgments}
Y.F. thanks Mar\'ia A. H. Vozmediano, Alberto Cortijo, and Jens H. Bardarson for encouragement and making this collaboration possible, and ICMT for its kind hospitality where part of this work was done. We thank Hart Goldman, Thors Hans Hansson, Anders Karlhede, Adolfo G. Grushin, and Flavio S. Nogueira for enlightening discussions. This work was supported in part by the ERC Starting Grant No. 679722. E.F. acknowledges support from the US National Science Foundation through grant No. DMR 1725401 at the University of Illinois.

\appendix

\section{Effective action manipulations}

\subsection{Effective action for the scalar field as a worldline path integral}
\label{sec:worldline}

To write the effective action, Eq. \eqref{eq primed effective action}, as an integral over worldlines, we  make use of the following identities
\beq
\partial_s H^{-s}|_{s\rightarrow0}=-\ln H,
\eeq
\beq
H^{-s}=\frac{\int^\infty_0 dt\,t^{s-1}e^{-Ht}}{\int_0^\infty dt\,t^{s-1}e^{-t}}=\frac{1}{\Gamma(s)}\int^\infty_0 dt\,t^{s-1}e^{-Ht},
\eeq
Using $\Gamma^{-1}(s)=s+\mathcal{O}(s^2)$, we can write
\beq
-\Gamma'=-\mathrm{tr}\ln H=\mathrm{tr}\int_{\epsilon}^\infty dt\,t^{-1}\,e^{-tH},
\eeq
with
\beq
H=-\frac{1}{\sqrt{g}}D_\mu(a,A)\sqrt{g}\,g^{\mu\nu}D_\nu(a,A)+\frac{R}{4}+m^2+2i\sqrt{\lambda}\,\sigma,
\eeq
and where we introduced a UV cut-off $\epsilon$. Now let us do time slicing and discretize the action as a sum over \textit{N} random paths of length $\Delta t$
\beq
-\Gamma=\int_{\epsilon}^\infty dt\,t^{-1}\int\prod_{i=1}^N d^3X_i\,\delta(X_N-X_0)\langle X_i|e^{-\Delta t\,H}|X_{i-1}\rangle,
\eeq
where $\Delta t=t/N$ and
$$
\langle X_i|e^{-\Delta t\,H}|X_{i-1}\rangle
$$
$$
=\int\frac{d^3 p}{(2\pi)^3}\exp\Big[-\Delta t\Big(ip_\mu\Delta X^\mu_i+\big(p_\mu+a_\mu+A_\mu\big)\big(p^\mu+a^\mu+A^\mu\big)+\frac{R}{4}+m^2+2i\sqrt{\lambda}\,\sigma\Big)\Big]
$$
\beq
=\sqrt{g}\,\left(\frac{1}{4\pi\Delta t}\right)^{3/2}\exp\Big[-\Delta t\Big(\frac{1}{4}g_{\mu\nu}\Delta X^\mu_i \Delta X^\nu_i-i(a_\mu+A_\mu)\Delta X^\mu_i+\frac{R}{4}+m^2+2i\sqrt{\lambda}\,\sigma\Big)\Big],
\eeq
with $\Delta X^\mu_i=(X^\mu_i-X^\mu_{i-1})/\Delta t$. Now we take the limit $\Delta t\rightarrow0$ and obtain
\beq
-\Gamma=\int_{\epsilon}^\infty dt\,t^{-1}\,DX^\mu\,\exp\left[-\int_0^tds\Big(\frac{1}{4}\dot{X}^2-i(a_\mu+A_\mu)\dot{X}^\mu+\frac{R}{4}+m^2+2i\sqrt{\lambda}\,\sigma\Big)\right],
\eeq
with $X^\mu(t)=X^\mu(0)$ and $\dot{X}^2=g_{\mu\nu}\dot{X}^\mu\dot{X}^\nu$. The integration measure is defined as
\beq
\int DX^\mu=\lim_{\Delta t\rightarrow0}\int\prod_{i=1}^N\left(\frac{1}{4\pi\Delta t}\right)^{3/2}\sqrt{g_i}\,d^3X_i.
\eeq
It fulfills the normalization condition
\beq
\int\prod_{i=1}^N\left(\frac{1}{4\pi\Delta t}\right)^{3/2}\sqrt{g_i}\,d^3X_i\,e^{-\Delta t\,g_{i,\mu\nu}\Delta X^\mu_i\Delta X^\nu_i/4}=1.
\eeq

\subsection{Effective action as a regularized path integral over the unit tangent vector}
\label{sec:path-integral-e}
Although not relevant for the main results of this work, we would like to show how one can further manipulate the worldline effective actions Eq.\eqref{eq:duality-boson} and \eqref{eq:duality-fermions}, and write them as a regularized path integral over the unit vector tangent to the worldlines, $e^{1\mu}$. We start from the following general effective action
\beq
-\Gamma=-\int_{\epsilon}^\infty dt\,t^{-1}\,DX^\mu\,\exp\left[-\int_0^tds\Big(\frac{1}{4}\dot{X}^2-iB_\mu\dot{X}^\mu+M^2\Big)\right]\,e^{i\Omega_{flat}/2},
\eeq
where $B_\mu$ and $M$ are generic vector and scalar fields, respectively. Let us regularize it by doing time slicing over \textit{N} random paths of length $\tilde{\epsilon}=\Delta t$, as done in appendix \ref{sec:worldline}
$$
-\Gamma=-\int_{\epsilon}^\infty dt\,t^{-1}\left(\frac{1}{4\pi\tilde{\epsilon}}\right)^{3N/2}\int\Bigg(\prod_{i=1}^N \sqrt{g_i}\,d^3X_i\,\delta(X_N-X_0)
$$
\beq
\times \,\exp\Big[-\tilde{\epsilon}\Big(\frac{1}{4}\Delta X^2_i-iB_\mu\Delta X^\mu_i+M^2\Big)\Big]\Bigg)\,e^{i\Omega_{flat}/2},
\eeq
where $\tilde{\epsilon}$ acts as an UV cut-off. 
Now one can do a change of integration variables $v^\mu_i=\Delta X^\mu_i$ and write the delta function as
\beq
\delta(X_N-X_0)=\delta(\tilde{\epsilon}\sum_{i=1}^N v_i)=\prod_{i=1}^N\int \frac{d^3k}{(2\pi)^3}\,e^{\tilde{\epsilon}ik_\mu v^\mu_i},
\eeq
so the effective action can be written as
$$
-\Gamma=-\int_{\epsilon}^\infty dt\,t^{-1}\left(\frac{\tilde{\epsilon}}{4\pi}\right)^{3N/2}\int \frac{d^3k}{(2\pi)^3}\int \Bigg(\prod_{i=1}^N\sqrt{g_i}\,\hat{v}^\mu_i\,dv_i\,v_i^2
$$
\beq
\times\, \exp\Big[-\tilde{\epsilon}\Big(\frac{1}{4}v_i^2-iv_i(k_\mu+B_\mu)\hat{v}^\mu_i+M^2\Big)\Big]\Bigg)\,e^{i\Omega_{flat}/2},
\eeq
where we have done $\int dv^\mu_i=\int d\hat{v}^\mu_i\,dv_i\,v_i^2$, and we should remember that $\Omega_{flat}$ does not depend on $v_i$, but only on the unit vector $\hat{v}^\mu_i$. Integrating over the radial coordinates $v_i$ and doing the change of variables $L=4t/\sqrt{\pi\tilde{\epsilon}}$ \cite{SSS90}
$$
-\Gamma=-\int_{\epsilon'}^\infty dL\,L^{-1}\int \frac{d^3k}{(2\pi)^3}\int \Bigg(\prod_{i=1}^N\sqrt{g_i}\,\frac{d\hat{v}^\mu_i}{4\pi}\,e^{-\frac{\sqrt{\pi\tilde{\epsilon}}}{4}M^2\Delta L}
$$
\beq
\times\,e^{\Delta L\,i(k_\mu+B_\mu)\hat{v}^\mu_i}\,e^{i\mathcal{O}(\sqrt{\tilde{\epsilon}})}\Bigg)\,e^{i\Omega_{flat}/2},
\eeq
where $\Delta L=L/N$, and
\beq
\mathcal{O}(\sqrt{\tilde{\epsilon}})\sim\sqrt{\tilde{\epsilon}}\,\Delta L\Big((k_\mu+B_\mu)\hat{v}^\mu_i\Big)^2+\mathcal{O}(\tilde{\epsilon}^{3/2})
\eeq
\beq
\epsilon'=\frac{4}{\sqrt{\pi\tilde{\epsilon}}}\epsilon.
\eeq
Taking the continuum limit $\Delta L\rightarrow0$ and renaming $e^{1\mu}\equiv\hat{v}^\mu$
\beq
-\Gamma=-\int \frac{d^3k}{(2\pi)^3}\int_{\epsilon'}^\infty \frac{dL}{L}\,e^{-\int_0^Lds\frac{\sqrt{\pi\tilde{\epsilon}}}{4}M^2}\int De^{1\mu}\,e^{i\int_0^Lds(k_\mu+B_\mu)e^{1\mu}}\,e^{i\Omega_{flat}/2}\,e^{i\mathcal{O}(\sqrt{\tilde{\epsilon}})},
\eeq
\beq
\mathcal{O}(\sqrt{\tilde{\epsilon}})\sim\sqrt{\tilde{\epsilon}}\int_0^Lds\Big((k_\mu+B_\mu)e^{1\mu}\Big)^2+\mathcal{O}(\tilde{\epsilon}^{3/2})
\eeq
with $De^{1\mu}=\delta(e^1-1)de^{1\mu}/4\pi=\prod_{i=1}^N \delta(e^1_i-1)\sqrt{g_i}\,de^{1\mu}_i/4\pi$.

Using the previous result, we can write the effective actions for the composite boson and the two time-reversal partner Dirac fermions, Eqs. \eqref{eq:duality-boson} and \eqref{eq:duality-fermions}, as
$$
\Gamma_{scalar+flux}=\int \frac{d^3k}{(2\pi)^3}\int_{\epsilon'}^\infty dL\,L^{-1}\,e^{-\int_0^Lds\,\frac{\sqrt{\pi\tilde{\epsilon}}}{4}\,\big(\frac{R}{4}+m^2+2i\sqrt{\lambda}\,\sigma\big)}
$$
\beq
\times\int De^{1\mu}\,\,e^{i\int_0^Lds\big(k_\mu+A_\mu-\frac{1}{4}e^{1a}\epsilon_{abc}\,\tensor{\omega}{_\mu^b^c}\big)e^{1\mu}}e^{\frac{i}{2}\Omega_{flat}}e^{i\mathcal{O}(\sqrt{\tilde{\epsilon}})}\,+iCS_g,
\eeq
with
\beq
\mathcal{O}\big(\sqrt{\tilde{\epsilon}}\big)\sim\sqrt{\tilde{\epsilon}}\int_0^Lds\,\Big[\big(k_\mu+A_\mu-\frac{1}{4}e^{1a}\epsilon_{abc}\,\tensor{\omega}{_\mu^b^c}\big)e^{1\mu}\Big]^2+\mathcal{O}\big(\tilde{\epsilon}^{3/2}\big),
\eeq
and
$$
2\mathcal{R}(\Gamma_f)=\int \frac{d^3k}{(2\pi)^3}\int_{\epsilon'}^\infty dL\,L^{-1}\,e^{-\int_0^Lds\,\frac{\sqrt{\pi\tilde{\epsilon}}}{4}\,\big(\frac{R}{4}+m^2+2i\sqrt{\lambda}\,\sigma\big)}
$$
\beq
\times\int De^{1\mu}\,\sum_{\alpha=\pm1}\,e^{i\int_0^Lds\,(k_\mu+A_\mu+\alpha iF^*_\mu(A)-\alpha\frac{1}{4}e^{1a}\epsilon_{abc}\,\tensor{\omega}{_\mu^b^c})e^{1\mu}}\,e^{\alpha\frac{i}{2}\Omega_{flat}}\,e^{i\mathcal{O}_\alpha\big(\sqrt{\tilde{\epsilon}}\big)},
\eeq
with
\beq
\mathcal{O}_\alpha\big(\sqrt{\tilde{\epsilon}}\big)\sim\sqrt{\tilde{\epsilon}}\int_0^Lds\,\Big[\big(k_\mu+A_\mu-\alpha\frac{1}{4}e^{1a}\epsilon_{abc}\,\tensor{\omega}{_\mu^b^c}\big)e^{1\mu}\Big]^2+\mathcal{O}\big(\tilde{\epsilon}^{3/2}\big).
\eeq

\section{The twist as a Berry phase}
\label{sec:twist-as-berry}

We  follow \cite{CL89} and show that the twist,
\beq
T=\frac{1}{2\pi}\int_0^t ds\,\bm{e}^1\cdot(\bm{e}^2\times\dot{\bm{e}}^2),
\eeq
can be written as a Berry phase \cite{CL89,TJ89,KK91}, which itself can be written as a Wess-Zumino-Witten term. Defining $\omega=\omega_1\bm{e}^1+\omega_i\bm{e}^i$ as the rate of rotation of the frame along \textit{C}: $\dot{\bm{e}}^1=\omega\times\bm{e}^1$, $\dot{\bm{e}}^i=\omega\times\bm{e}^i$ ($i=1,2$), we can rewrite \textit{T} as
\beq
T=\frac{1}{2\pi}\int_0^t \,\omega_1ds,
\label{eq:twistomega}
\eeq
with $\omega_1=\dot{\bm{e}}^2\times\bm{e}^3$.

Let us now introduce two $CP^1$ fields \textit{z} and \textit{w} which satisfy
\beq
z^\dag\bm\sigma z=\bm{e}^1,\quad w^\dag\bm\sigma w=-\bm{e}^1,
\label{Eq. definition cp1}
\eeq
so that $w_1=-z_2^*$, $w_2=z_1^*$. Next we define the complex vector $\bm{v}=z^\dag\bm\sigma w$ which, using Eq. (\ref{Eq. definition cp1}), is easy to show to be orthogonal to $\bm{e}^1$. By further using the properties of the sigma matrices, one can prove that $\bm{v}$ is equal to the combination  $\bm{e}^2-i\bm{e}^3$ up to a phase  \cite{CL89}
\beq
\bm{v}=e^{i\alpha}(\bm{e}^2-i\bm{e}^3)\label{eq uptoaphase},
\eeq
with $\alpha(t)=\alpha(0)+2\pi \nu$, where $\nu\in\mathbb{Z}$ is the winding number. Recalling the definition $\omega_1=\dot{\bm{e}}^2\times\bm{e}^3$, one can show that
\beq
\bm{v}\cdot\dot{\bm{v}}^*=4z^\dag\cdot\dot{z}=-2i(\dot{\alpha}+\omega_1).
\eeq
We get
\beq
T=\frac{1}{2\pi}\int_0^t \omega_1\, ds=\frac{i}{\pi}\int_0^t\, z^\dag\cdot\dot{z}-\frac{1}{2\pi}\int_0^t\dot{\alpha}\, ds=-\frac{1}{\pi}\int_0^t A(s)\, ds-\nu,
\eeq
where we have introduced the Berry connection $A=-iz^\dag\cdot\dot{z}$, so that $T$ is written as a Berry phase plus a winding number. Let us study the dependence of the twist on the framing, by performing the rotation
\beq
\tilde{\bm e}^2=\cos\varphi\,\bm e^2+\sin\varphi\,\bm e^3,\quad\quad \tilde{\bm e}^3=-\sin\varphi\,\bm e^2+\cos\varphi\,\bm e^3,
\eeq
where periodicity demands $\varphi(t)=\varphi(0)+2\pi m$, with $m\in\mathbb{Z}$. This shifts the phase in Eq. (\ref{eq uptoaphase}): $\tilde{\alpha}=\alpha+\varphi$. Therefore, the twist transforms as
\beq
\tilde{T}=T-m,
\eeq
which confirms its frame dependence.

Finally, as we are dealing with closed paths, the Berry phase is just the enclosed flux of the Berry curvature, which can be defined by continuing $s\in C$ to $u\in S$, where \textit{S} is the surface enclosed by \textit{C}
\beq
\frac{1}{\pi}\int_0^t A(s)\, ds=\frac{1}{\pi}\int_S d^2u(\partial_{u_1} A_2-\partial_{u_2} A_1)=\frac{1}{2\pi}\int_0^1 d\rho\int_0^t ds\, \bm{e}^1\cdot(\partial_{\rho} \bm{e}^1\times\partial_{s} \bm{e}^1).
\eeq
In the last equality the surface $S$ is parametrized by $s\in[0,t]$ and $\rho\in[0,1]$, and $\bm{e}^1$ is extended to $\bm{e}^1(t,\rho)$ in such a way that $\bm{e}^1(t,0)=\bm{e}_0$ and $\bm{e}^1(t,1)=\bm{e}^1(t)$, with $\bm{e}_0$ some constant vector. The last term is precisely the Wess-Zumino-Witten term \cite{WZ71,WW-Z83}, and is equal to the signed area enclosed by the path \textit{C} on the sphere $S_2$, defined by $\bm{e}^1: C\rightarrow S_2$. Then, the gauge field $A(s)$ can be interpreted as that of a monopole at the center of the sphere, so that the signed area is proportional to the magnetic flux enclosed.

\section{The spin factor as a path integral over anticommuting fields}
\label{sec:spin-as-pathintegral}

Let us show how, in the absence of curvature, the spin factor can be defined as a path integral over one dimensional fermions, and how one can then extend this definition to curved space. We start with the following path integral over anticommuting fields $\psi^a$ \cite{P88LH,GHK90}
\beq
\int D\psi^aD\chi\,e^{\int_0^t ds\,(\frac{1}{4}\psi_a\dot{\psi}_a+\chi e^1_a\psi_a)},
\label{eq:pathfermiononed1}
\eeq
where $\chi$ acts as a Lagrange multiplier imposing transversality of the fields $\psi^a$, as can be seeing by integrating it out
\beq
\int D\psi^a\,\delta(\psi^ae^1_a)\,e^{\frac{1}{4}\int_0^t ds\,\psi_a\dot{\psi}_a}.
\eeq
Doing the change of variables
\beq
\psi_a=2e^1_a\phi^1+2e^i_a\phi_i,
\eeq
with $i=2,3$, we get
\beq
\int D\phi^1D\phi^i\,\delta(2\phi^1)\,e^{\int_0^t ds\,(\phi_i\dot{\phi}_i+\omega_1\epsilon_{ij}\phi_i\phi_j)},
\eeq
where $\omega_1$ is defined in appendix \ref{sec:twist-as-berry}. Taking into account that the correlator $\langle\phi_i(s)\phi_j(s')\rangle$ with respect to the action $\int_0^tds\,\phi_i\dot{\phi}_i$ is $\delta_{ij}\,sgn(s-s')$, it is not difficult to prove that \cite{P88LH,GHK90}
\beq
\int D\phi^1D\phi^i\,\delta(2\phi^1)\,e^{\int_0^t ds\,(\phi_i\dot{\phi}_i+\omega_1\epsilon_{ij}\phi_i\phi_j)}=\frac{\mathrm{\tilde{t}r}}{2}\,\mathcal{P}\,e^{\frac{1}{4}\int_0^tds\,\omega_1\epsilon_{ij}\sigma_i\sigma_j},
\label{eq integral sigmas}
\eeq
where $\mathcal{P}$ denotes path ordering and $\mathrm{\tilde{t}r}$ is the trace over spin indexes. We get
\beq
\int D\psi^aD\chi\,e^{(\frac{1}{4}\int_0^t ds\,\psi_a\dot{\psi}_a+\chi e^1_a\psi_a)}=\frac{\mathrm{\tilde{t}r}}{2}\,\mathcal{P}\,e^{\frac{1}{4}\int_0^tds\,\omega_1\epsilon_{ij}\sigma_i\sigma_j}=\frac{\mathrm{\tilde{t}r}}{2}e^{\frac{i}{2}\int_0^tds\,\omega_1\sigma_1}=(-1)^\nu\cos{\Omega_{flat}/2}.
\eeq
For the last equality we used the relation $SL=W+T$ together with eqs. \eqref{eq:twistomega} and \eqref{eq writhe 1}, which give expressions for the twist and writhe in terms of $\omega_1$ and $\Omega_{flat}$, respectively.

Let us write the above result as
\beq
\cos{\Omega_{flat}/2}=(-1)^\nu\int D\phi^i\,e^{\int_0^t ds\,(\phi^T\dot{\phi}+i\omega_1\phi^T\sigma_2\,\phi)},
\label{eq integral gauge invariance}
\eeq
with $\phi^T=(\phi_2,\phi_3)$. There is a subtlety here. It is apparent that the action in the r.h.s. is gauge invariant under
\beq
\phi\rightarrow e^{-i\alpha(s)\sigma_2}\,\phi,
\eeq
\beq
i\omega_1\sigma_2\rightarrow e^{-i\alpha(s)\sigma_2}\, i\omega_1\sigma_2\,e^{i\alpha(s)\sigma_2}+e^{-i\alpha(s)\sigma_2}\,\partial_s\,e^{i\alpha(s)\sigma_2}\,\Longrightarrow\, \omega_1\rightarrow\omega_1+\dot{\alpha}(s),
\eeq
where $\mathrm{exp}[-i\alpha(s)\sigma_2]$ is a periodic function in $[0,t]$. On the other hand, the l.h.s. is only invariant under small gauge transformations, i. e. transformations that are continuously connected to the identity. In fact, large gauge transformations of the form $\alpha(s)=2n\pi s/t$, with $n\in\mathbb{Z}$, transform the l.h.s. of (\ref{eq integral gauge invariance}) as
\beq
\cos{\Omega_{flat}/2}\rightarrow(-1)^n\cos{\Omega_{flat}/2},
\eeq
so that large gauge invariance is lost. This means that the path integral in the r.h.s. of Eq. (\ref{eq integral gauge invariance}) is gauge anomalous, i.e. it can not be regularized in a gauge invariant way \cite{ERFS86}.

Take now the result for the spin factor given by integral \eqref{eq:pathfermiononed1}
\beq
\cos{\Omega_{flat}/2}=(-1)^\nu\int D\psi^aD\chi\,e^{\int_0^t ds\,(\frac{1}{4}\psi_a\dot{\psi}_a+\chi e^1_a\psi_a)},
\eeq
From this expression it is straight forward to introduce curvature, as was done in  \cite{GHK90}
\beq
\cos{\Omega/2}=(-1)^\nu\int D\psi^\mu D\chi\,e^{\frac{1}{4}\int_0^t ds\,g_{\mu\nu}\psi^\mu \dot{X}^{\gamma}\nabla_\gamma\psi^\nu }\,e^{\int_0^t ds\,\chi\, g_{\mu\nu}e^{1\nu}\psi^\mu},
\label{eq cosine curved 1}
\eeq
where $D\psi^\mu=\prod_i\sqrt{g(X_i)}d\psi^\mu(X_i)$, and $\dot{X}^{\gamma}\nabla_\gamma$ is the directional covariant derivative along \textit{C}, with $\nabla_\gamma\psi^\nu=\partial_\gamma\psi^\nu+\Gamma^\nu_{\beta\gamma}\psi^\beta$. Using the tetrad postulate
\beq
\nabla_\gamma E^\nu_a=\partial_\gamma E^\nu_a+\Gamma^\nu_{\beta\gamma}E^\beta_a+\tensor{\omega}{_\gamma_a^b}E^\nu_b=0,
\eeq
with $E^\nu_a$ the tetrad and $\tensor{\omega}{_\gamma_a^b}$ the spin connection, we can write
\beq
\dot{X}^{\gamma}\nabla_\gamma\psi^\nu=\dot{X}^{\gamma}\nabla_\gamma(E^\nu_a\psi^a)=\dot{\psi}^\nu+\psi^a\dot{X}^{\gamma}(\partial_\gamma E^\nu_a+\Gamma^\nu_{\beta\gamma}E^\beta_a)=\dot{\psi}^\nu-\psi^a\dot{X}^{\gamma}\tensor{\omega}{_\gamma_a^b}E^\nu_b.
\eeq
We then get
\beq
\cos{\Omega/2}=(-1)^\nu\int D\psi^a D\chi\,e^{\frac{1}{4}\int_0^t ds\,\psi_a\dot{\psi}^a}\,e^{\int_0^t ds\,(\chi e^{1\mu}\psi_\mu+\frac{1}{4}\dot{X}^\mu\tensor{\omega}{_\mu^a^b}\psi_a\psi_b)},
\eeq
where we have done
\beq
D\psi^\mu=\prod_i\sqrt{g(X_i)}\,d\psi^\mu(X_i)=\prod_i\sqrt{g(X_i)}\,\frac{d\psi^a(X_i)}{\sqrt{g(X_i)}}=\prod_id\psi^a(X_i)=D\psi^a.
\eeq
We integrate out $\psi^a$ as we did to obtain (\ref{eq integral sigmas})
\beq
\cos{\Omega/2}=(-1)^\nu\frac{\mathrm{\tilde{t}r}}{2}\,\mathcal{P}\,e^{\frac{i}{2}\int_0^t ds\,\omega_1\sigma_1}\,e^{\frac{1}{8}\int_0^t ds\,\dot{X}^{\mu}\tensor{\omega}{_\mu^a^b}[\sigma_i,\sigma_j]e^i_ae^j_b}.
\eeq
Doing
\beq
[\sigma_i,\sigma_j]e^i_ae^j_b=2i\epsilon_{ij}e^i_ae^j_b\sigma_1=2ie^{1c}\epsilon_{cab}
\eeq
we can express the exponential as
\beq
e^{\frac{i}{2}\int_0^t ds\,(\omega_1+\frac{1}{2}\dot{X}^{\mu}e^{1a}\epsilon_{abc}\tensor{\omega}{_\mu^b^c})\sigma_1}.
\eeq
Computing the trace we obtain
\beq
\cos{\frac{\Omega}{2}}=\cos\big(\frac{1}{2}\int_0^t ds\,\omega_1+\frac{1}{4}\int_0^t ds\,\dot{X}^{\mu}e^{1a}\epsilon_{abc}\tensor{\omega}{_\mu^b^c}+\pi\nu\big),
\eeq
which in terms of $\Omega_{flat}$ is
\beq
\cos{\Omega/2}=\cos{\left(\Omega_{flat}/2-\frac{1}{4}\int ds\,\dot{X}^{\mu}e^{1a}\epsilon_{abc}\,\tensor{\omega}{_\mu^a^b}\right)}.
\label{eq cosine curved 2}
\eeq





\bibliography{bibbosonfermionduality}




\end{document}